%% file: ApJ.tex
\documentclass[twocolumn,tighten,times]{aastex63}
\usepackage{graphicx}
\usepackage{subfigure}
\usepackage{amsmath}
\usepackage{amssymb}
\usepackage{natbib}
\usepackage{sidecap}
\usepackage{cancel}
\usepackage{multirow}
\usepackage{stmaryrd}
\input{aas_macros}

\newcommand{\be}{\begin{equation}}
\newcommand{\ee}{\end{equation}}

\begin{document}

\title{High-Energy Neutrinos from Gamma-Ray-Faint Accretion-Powered Hypernebulae}

\author[0000-0002-5519-9550]{Navin Sridhar}
\affiliation{Department of Astronomy, Columbia University, New York, NY 10027, USA}
\affiliation{Theoretical High Energy Astrophysics (THEA) Group, Columbia University, New York, NY 10027, USA}
\correspondingauthor{Navin Sridhar}
\email{navin.sridhar@columbia.edu}

\author[0000-0002-4670-7509]{Brian D. Metzger}
\affiliation{Theoretical High Energy Astrophysics (THEA) Group, Columbia University, New York, NY 10027, USA}
\affiliation{Department of Physics, Columbia University, New York, NY 10027, USA}
\affiliation{Center for Computational Astrophysics, Flatiron Institute, New York, New York 10010, USA}

\author[0000-0002-5387-8138]{Ke Fang}
\affiliation{Department of Physics, Wisconsin IceCube Particle Astrophysics Center, University of Wisconsin, Madison, WI, 53706, USA}

\begin{abstract}
Hypernebulae are inflated by accretion-powered winds accompanying hyper-Eddington mass transfer from an evolved post-main sequence star onto a black hole or neutron star companion. The ions accelerated at the termination shock---where the collimated fast disk winds/jet collide with the slower, wide-angled wind-fed shell---can generate high-energy neutrinos via hadronic ($pp$) reactions, and photohadronic ($p\gamma$) interactions with the disk thermal and Comptonized nonthermal background photons.  It has been suggested that some fast radio bursts (FRBs) may be powered by such short-lived jetted hyper-accreting engines. Although neutrino emission associated with the ms-duration bursts themselves is challenging to detect, the persistent radio counterparts of some FRB sources---if associated with hypernebulae---could contribute to the high energy neutrino diffuse background flux.  If the hypernebula birth rate follows that of steller-merger transients and common envelope events, we find that their volume-integrated neutrino emission---depending on the population-averaged mass-transfer rates---could explain up to $\sim25\%$ of the high-energy diffuse neutrino flux observed by the IceCube Observatory and the Baikal-Gigaton Volume Detector (GVD) Telescope.  The time-averaged neutrino spectrum from hypernebula---depending on the population parameters---can also reproduce the observed diffuse neutrino spectrum. The neutrino emission could in some cases furthermore extend to $>$100\,PeV, detectable by future ultra-high-energy neutrino observatories. The large optical depth through the nebula to Breit-Wheeler ($\gamma\gamma$) interaction attenuates the escape of GeV-PeV gamma-rays co-produced with the neutrinos, rendering these gamma-ray-faint neutrino sources, consistent with the \textit{Fermi} observations of the isotropic gamma-ray background. 
\end{abstract}

\keywords{Neutrino Astronomy (1100); Radio transient sources (2008); Ultraluminous X-ray sources (2164); X-ray binary stars (1811); Shocks (2086); Plasma astrophysics (1261); High energy astrophysics (739)}

\section{Introduction}
 
The IceCube Observatory, and recently, the Baikal-GVD Telescope measured the flux and spectrum of the diffuse high-energy (TeV-100\,PeV; HE, hereafter) neutrino background, finding an all-flavor all-sky-averaged flux of $\sim5\times10^{-8}\,{\rm GeV\,s^{-1}\,cm^{-2}\,sr^{-1}}$ at 100~TeV \citep{Aartsen+18, Aartsen+20a, Abbasi+21, Abbasi+22a, Baikal_Collaboration22}. The sensitivity limits of the current neutrino facilities have not allowed for the detection of the long sought-after ultra-high-energy ($\ge$100\,PeV; UHE, hereafter) neutrinos. The detected HE neutrinos may plausibly be produced in baryonic shocks: one of the consequences of magnetized shocks of electron-ion plasmas is the non-thermal acceleration of ions to relativistic energies via diffusive shock acceleration (\citealt{Blandford&Ostriker78}; and also via turbulence, for example, \citealt{Comisso&Sironi22}). Such relativistic ions can generate neutrinos via hadronuclear ($pp$), and photohadronic ($p\gamma$) interactions (e.g., \citealp{Waxman&Bahcall97, Meszaros&Waxman01, Dermer&Atoyan03, Guetta&Granot03}). The contribution from the Galactic plane to the all-sky neutrino flux is found to be $\lesssim$5\%-10\% at around 100\,TeV, suggesting that most of the observed apparently isotropic diffuse neutrino flux is extragalactic \citep{IceCube:2017trr, ANTARES:2018nyb, Fang&Murase21}. There has been evidence for at least one gamma-ray blazar, TXS\,0506+056, contributing to the neutrino signal \citep{IceCube_Blazar1+18, IceCube_Blazar2+18}. Recently, the ten-year point-source searches with IceCube have indicated that the type-II Seyfert starburst galaxy NGC\,1068 is the most significant (at $4.2\sigma$) steady source of neutrinos \citep{Aartsen+20, IceCubeCollaboration_22}. However, in their respective energy ranges, NGC\,1068 and TXS\,0506+056 contribute no more than $\sim$1\% above 10\,TeV to the overall HE diffuse background neutrino flux.

Therefore, the sources for the majority of the extragalactic background neutrinos still remain a mystery.  Various astrophysical objects have been studied as the sources of the HE neutrinos, such as gamma-ray bursts \citep[GRBs;][]{Paczynski&Hu94, Waxman&Bahcall97, Meszaros&Waxman01, Dermer&Atoyan03, Fasano+21, Kimura_22}, active galactic nuclei \citep[AGN;][]{Eichler79, Stecker+91, Jacobsen+15, Murase17, Kun+22}, different kinds of supernovae \citep[SNe;][]{Razzaque+04, Murase+11, Murase18, Grichener&Soker21, Chang+22, Sarmah+22, Abbasi+23}, bubbles associated with ultraluminous X-ray (ULX) sources \citep{Inoue+17}, starburst galaxies \citep{Romero&Torres03, Loeb&Waxman06, Thompson+06}, tidal disruption events \citep[TDEs;][]{Wang&Liu16, Dai&Fang17, Senno+15, Senno+17, Lunardini&Winter17}, nonrelativistic shock-powered transients \citep[such as e.g., luminous red novae, fast blue optical transients;][]{Fang+20}, fast radio bursts \citep[FRBs;][]{Metzger+20}, proto-magnetars \citep{Bhattacharya+22}, supermassive black hole mergers \citep{Jaroschewski+22}, collapsars \citep{Guo+22}, among other sources.  However, stacking analysis with IceCube and ANTARES has shown that none of the above sources are likely to contribute a major fraction of the all-sky neutrino flux \citep{Aartsen+16, Albert+21, Liodakis+22}. 

The HE diffuse background neutrino flux is comparable to that of UHE cosmic rays and gamma-rays, hinting at a possible common origin. This is expected because a comparable flux of gamma-rays will also be produced with neutrinos following both the proton-proton ($pp$) and proton-photon ($p\gamma$) interactions. But, the diffuse isotropic gamma-ray background (IGRB; subtracting the contributions from resolved sources, foreground emission, and unresolved blazars) observed by \textit{Fermi} Large Area Telescope (LAT) between 100\,MeV and 1\,TeV \citep{Ackermann+15} has disfavored gamma-ray bright sources being dominant sources of HE neutrinos, i.e., at lower energies, there is more neutrino flux than the corresponding gamma-ray flux if they were produced by the same sources through the $pp$ or $p\gamma$ channels \citep{IceCube21}. This fact that the all-sky neutrino flux at 10\,TeV is larger than the expected gamma-ray flux (from an accelerated proton with a powerlaw index of $-2$) suggests that the dominant source of the HE diffuse background neutrino flux must be opaque to GeV-TeV gamma-rays \citep{Murase+16, Capanema+20, Capanema+21, Fang+22}.  

Such requirements could naturally be met by the new class of astrophysical objects called `hypernebulae' \citep{Sridhar&Metzger22}. A hypernebula is an energetic, compact ``bubble'' ($\lesssim1$\,pc; much smaller and powerful than the nebulae surrounding typical ULX sources) of plasma inflated into the circumbinary medium through shocks by powerful accretion disk winds and jets during the dynamically unstable (or even thermal-timescale) transfer of matter onto a compact object (black hole or neutron star) at rates that exceed the Eddington rate by many orders of magnitude. Such brief phases of rapid mass transfer are expected to occur following the Roche-lobe overflow from a donor star, prior to a common-envelope merger event and the `intermediate luminosity optical transients' accompanying it (such as Luminous Red Novae; \citealt{Soker22}, and Fast Blue Optical Transients \citealt{Prentice+18, Perley+19, Ho+20}). The compact object, upon entering the common envelope, can continue accreting matter from the stellar envelope and launch jets, where accelerated protons can generate neutrinos via photohadronic interactions \citep{Grichener&Soker21}. Recently, \cite{Sridhar+21b} showed that the flaring collimated jets launched by the compact object at the heart of such accreting engines could power the (periodically) repeating FRBs. The radio-bright hypernebulae surrounding them could be the source of the persistent radio emission seen spatially coincident with FRBs \citep{Chatterjee+17, Niu+22}, which also imparts the rotation measure onto the pulses propagating through them \citep{Sridhar&Metzger22}.  In fact, hypernebulae could already be present in blind wide-field radio surveys such as FIRST \citep{Becker+95} and VLASS \citep{Lacy+20} as off-nuclear radio sources, and could be spatially resolved by the next generation VLA \citep{Selina+18}.

In this paper, we investigate the properties of the ions accelerated at the different shock fronts present in hypernebulae, and their interactions with other ions ($pp$) and the `background' disk thermal and Comptonized nonthermal UV/X-ray photons ($p\gamma$). Such interactions could produce pions which subsequently decay into neutrinos and gamma-rays.  We show that the HE neutrinos generated by hypernebulae could explain---depending on the mass-transfer rates achieved in typical sources---a significant fraction ($\gtrsim 25\%$) of the observed extragalactic HE diffuse neutrino background.  The dense environment present in hypernebulae further attenuates the $>$TeV gamma-rays produced by pion decay, thus rendering them gamma-ray faint. In \S\ref{sec:engine_properties}, we discuss the nature of the outflows from the accreting engine: properties of the shocks and their associated timescales (\S\ref{subsec:disk_winds}), and the sources of background radiation (\S\ref{subsec:background_radiation}). The model for neutrino production is presented in \S\ref{sec:neutrino_model}, with discussions on pion production channels (\S\ref{subsec:pion_production}), pion production efficiency (\S\ref{subsec:interaction_timescales}), ion heating and acceleration (\S\ref{subsec:proton_energization}), and neutrino production from pion decay (\S\ref{subsec:neutrino_production}). We present some observational implications of our model in \S\ref{sec:observable_implications} before concluding in \S\ref{sec:conclusion}.

\section{Properties of the engine} \label{sec:engine_properties}

As discussed in \cite{Sridhar&Metzger22}, at least three different types of disk/jet outflow interactions can occur in hypernebulae: (1) The slow, wide-angled winds from the black hole/neutron star accretion disk interacting with the circumstellar medium (CSM) via a sub-relativistic forward shock (FS); the location of this shell---that sweeps up both the wind material and the CSM---controls the overall radius of the system, and hence, the gas and X-ray photon number density within the hypernebula. (2) The trailing slow winds colliding with the accumulated shell ahead of them, forming what we call the `wind termination shock' (WTS). (3) Faster winds/jet from the innermost regions of the accretion flow, directed along the instantaneous (and likely precessing) angular momentum axis of the inner disk/jet, inflate bipolar lobes upon their interaction with the swept-up shell, mediated by what we call the `jet termination shock' (JTS).  We refer the reader to Fig.~1 of \cite{Sridhar&Metzger22} for a schematic diagram of the hypernebula and these various interaction fronts.

Our assumed fiducial model parameters, corresponding to a high-mass X-ray binary, are as follows [with the explored parameter range given in square brackets]. The central compact object is fiducially assumed to be a black hole of mass $m_{\bullet}=M_{\bullet}/M_{\odot}=10$, experiencing mass-transfer at a rate of $\dot{m}=\dot{M}/\dot{M}_{\rm Edd}=10^5~[10^3,10^7]$ \footnote{$\dot{M}_{\rm Edd}=L_{\rm Edd}/(0.1c^2)$ is the Eddington accretion rate, and $L_{\rm Edd}\simeq 1.5\times10^{39}(m_\bullet/10)\,{\rm erg\,s^{-1}}$} from an evolved companion star of mass $m_{\star} = M_{\star}/M_{\odot} = 30$.  The accretion disk drives a wide-angled ``slow'' wind of average velocity $v_{\rm w}/c=0.03~[0.01,0.1]$ and kinetic luminosity $L_{\rm w}=\dot{M}_{\rm w}v_{\rm w}^2/2$, as well as a ``fast'' wind/jet from the innermost regions of the disk with a speed $v_{\rm j}/c=0.3~[0.1,0.5]$ \citep{Kobayashi+18, Urquhart+18, Pinto&Kosec22}.  The mass loss rate in the wide angled disk winds is assumed to equal the mass-transfer rate $\dot{M}_{\rm w}\approx \dot{m}$ \citep{Hashizume+15}, while that in the jet is taken to be $\dot{M}_{\rm j}=\eta\dot{M}_{\rm w}(v_{\rm w}/v_{\rm j})^2$, with $\eta = 0.5~[0.1,1]$ expected from radiatively-inefficient super-Eddington accretion disk models, due to the roughly equal gravitational energy released per radial decade in the accretion flow \citep{Blandford&Begelman99}.  We define the jet magnetization parameter $\sigma_{\rm j}=0.1~[0.01,1]$ as the ratio of its Poynting to kinetic energy flux.  The density of the external CSM with which the wind-shell interacts is $\rho_{\rm csm}=\mu n m_{\rm p}$, where $n=10\,[1,100]\,{\rm cm}^{-3}$, typical of the massive-star environments \citep{Abolmasov+08, Pakull+10, Toala&Arthur_2011}, and $\mu=1.38$ is the mean atomic weight.

\subsection{Ejecta} \label{subsec:disk_winds}

Mass-transfer from a donor star that has evolved off the main sequence can under some circumstances become unstable, ultimately culminating in a merger or common-envelope event (e.g., \citealt{Roepke&DeMarco22}).  However, prior to this final phase, the compact object can experience a high mass-transfer rate $\dot{M}$, lasting for a duration up to
\be\label{eq:t_active}
t_{\rm active}\sim\frac{M_\star}{\dot{M}}\approx 10^3\,{\rm yr}\left(\frac{m_\star}{30}\right)\dot{m}_{5}^{-1},
\ee
where $\dot{m}_{5} \equiv \dot{M}/(10^{5}\dot{M}_{\rm Edd})$.  This ``active time'', which precedes the final dynamical phase, corresponds to the lifetime of the jet inflating the hypernebula.  The ejecta freely expands until the mass released in the accretion disk-winds becomes comparable to that accumulated from the CSM, on the timescale \citep{Sridhar&Metzger22},
\be \label{eq:t_free}
t_{\rm free} \approx 80\,{\rm yr}\,\left(\frac{L_{\rm w,42}}{n_{1}}\right)^{1/2}v_{\rm w,9}^{-5/2}.
\ee
Unless otherwise specified, we shall normalize analytical estimates to $t=t_{\rm free}$ throughout the rest of the paper, using the shorthand notation $Y_{\rm x} \equiv Y/10^{\rm x}$ in cgs units for quantities other than $\dot{m}$.  

At times $t \gtrsim t_{\rm free}$, the nebula enters a self-similar deceleration phase \citep{Weaver+77}.  The temporal evolution of the radius of the shell---bridging the two phases across $t_{\rm free}$---can be approximated as,
\begin{equation}\label{eq:R_fs}
    R(t) \simeq
    \begin{cases}
          v_{\rm w}t \approx 0.75\,{\rm pc}\,v_{\rm w,9}\left(\frac{t}{80\,{\rm yr}}\right) & (t < t_{\rm free}) \\
          \alpha\left(\frac{L_{\rm w}t^{3}}{\rho_{\rm csm}}\right)^{1/5} \approx 0.75\,{\rm pc} \,\left(\frac{L_{\rm w,42}}{n_{1}}\right)^{1/5}\left(\frac{t}{80\,{\rm yr}}\right)^{3/5} & (t > t_{\rm free}),\\
    \end{cases}
\end{equation}
where $\alpha \approx 0.88$ while the forward shock is adiabatic and $\alpha \approx 0.76$ after it becomes radiative \citep{Weaver+77}\footnote{As we show below, for the cases of interest $\dot{m}>10^3$, the FS does not become radiative  on the active duration.}. The shell expands at roughly the forward shock velocity
\begin{equation}\label{eq:v_fs}
    v_{\rm fs} = \frac{dR}{dt}\approx 10^9\,{\rm cm\,s}^{-1}\times
    \begin{cases}
          v_{\rm w,9} & (t < t_{\rm free}) \\
           \left(\frac{L_{\rm w,42}}{n_{1}}\right)^{1/5}\left(\frac{t}{80\,{\rm yr}}\right)^{-2/5}& (t > t_{\rm free}),\\
    \end{cases}
\end{equation}
with the corresponding expansion timescale,
\be \label{eq:t_exp}
t_{\rm exp} \simeq \frac{R}{ v_{\rm fs}} \simeq 80\,{\rm yr} \left(\frac{t}{80\,{\rm yr}}\right).
\ee
As the shell expands, it accumulates a mass
\begin{equation} \label{eq:Msh}
 M_{\rm sh} \simeq 
    \begin{cases}
          \dot{M}_{\rm w}t \approx 2\,M_{\odot} \dot{m}_5\left(\frac{t}{80\,{\rm yr}}\right) & (t < t_{\rm free}) \\
          (4\pi/3)\rho_{\rm csm}R^{3} \approx 2\,M_{\odot} L_{\rm w,42}^{3/5}n_1^{2/5}\left(\frac{t}{\rm 80\,yr}\right)^{9/5} & (t > t_{\rm free}).
    \end{cases}
\end{equation}
At later times ($t>t_{\rm free}$), the swept-up mass from the CSM grows rapidly as $t^{9/5}$, dominating the mass injected by the disk wind.  The mean baryon density of the shell is given by,
\begin{align} \label{eq:n_p}
n_{\rm p}\simeq \frac{3M_{\rm sh}}{4\pi R^2 m_{\rm p}\mu {\rm d}R}
\approx
    \begin{cases}
          30\,{\rm cm}^{-3}\,\dot{m}_{5}v_{\rm w}^{-3}\left(\frac{t}{80\,{\rm yr}}\right)^{-2} & (t < t_{\rm free})\\
          10\,{\rm cm}^{-3}\,n_1 & (t > t_{\rm free}) \\
    \end{cases}
    ,
\end{align}
where we have assumed shell radial thickness $dR\sim R$, as expected for an adiabatic shock in which the post-shock gas does not cool rapidly (see below). This density is substantially lower, for example, than in a magnetar wind nebula \citep{Fang&Metzger17} during its active neutrino production phase ($\lesssim1$\,yr after magnetar birth), with implications for the interaction channels through which neutrinos are produced (see \S\ref{subsec:interaction_timescales} for further discussion).

The kinetic luminosity of the different shock fronts are approximately given by,
\be \label{eq:L_s}
    L_{\rm s} \approx \frac{9\pi}{8} R^2 \times
        \begin{cases}
            \rho_{\rm fs} v_{\rm fs}^3 &  ({\rm FS}) \\ 
            \rho_{\rm wts} (v_{\rm w} - v_{\rm fs})^3  &  ({\rm WTS}) \\
            \rho_{\rm jts} (v_{\rm j} - v_{\rm fs})^3  &  ({\rm JTS})
        \end{cases}
\ee
where the upstream gas densities entering the above expressions are $\rho_{\rm fs}=\rho_{\rm csm}$, $\rho_{\rm wts}=\dot{M}_{\rm w}/(4\pi v_{\rm w}R^2)$, and $\rho_{\rm jts}=\dot{M}_{\rm j}/(4\pi v_{\rm j}R^2)$, respectively. At $t>t_{\rm free}$, the particles behind each shock are heated to mean energies,
\be \label{eq:mean_temp}
\bar{E} \approx \frac{3}{16}\mu m_{\rm p} \times
    \begin{cases}
        v_{\rm fs}^2  \approx 220\,{\rm keV}\,\left(\frac{L_{\rm w,42}}{n_{1}}\right)^{2/5}\left(\frac{t}{\rm 80\,yr}\right)^{-4/5} &  ({\rm FS}) \\ 
        (v_{\rm w} - v_{\rm fs})^2 \approx 120\,{\rm keV}\,v_{\rm w,9}^2 &  ({\rm WTS}) \\
        (v_{\rm j} - v_{\rm fs})^2 \approx 17\,{\rm MeV}\,\left(\frac{v_{\rm j}}{0.3\,c}\right)^2 &  ({\rm JTS})
    \end{cases}
\ee
The corresponding cooling timescale of the post-shock thermal plasma is given by (e.g., \citealt{Vlasov+16}),
\be \label{eq:t_cool}
t_{\rm cool} \simeq \frac{3\bar{E}}{2n_{\rm s}\mu\Lambda[T_{\rm s}]}
\ee
where $n_{\rm s}=4\rho_{\rm s}/\mu m_{\rm p}$ is the post-shock density assuming strong shocks and an adiabatic index $\hat{\gamma}=5/3$.  We adopt a cooling function appropriate for solar-metallicity gas which includes both free-free emission and atomic lines $\Lambda[T_{\rm s}] = \Lambda_{\rm ff}[T_{\rm s}] + \Lambda_{\rm line}[T_{\rm s}]$, where \citep{Schure+09,Draine11},
\begin{subequations} \label{eq:cooling_function}
\begin{gather}
\Lambda_{\rm ff}[T_{\rm s}] \simeq 3\times10^{-27}(T_{\rm s}/{\rm K})^{0.5}\,{\rm erg~s}^{-1}{\rm cm}^{3},\\
\Lambda_{\rm lines}[T_{\rm s}] \simeq 2.8\times 10^{-18}(T_{\rm s}/{\rm K})^{-0.7}\,{\rm erg~s}^{-1}{\rm cm}^{3}.
\end{gather}
\end{subequations}
The radiative efficiency of the shocks can be quantified by the parameter \citep{Vlasov+16, Sridhar&Metzger22},
\be \label{eq:f_rad}
f_{\rm rad} = \left(1 + \frac{5}{2}\frac{t_{\rm cool}}{t_{\rm exp}}\right)^{-1},
\ee
where $f_{\rm rad} \ll 1 (\simeq 1)$ defines an adiabatic(radiative) shock, respectively.

\begin{figure} 
\centering
\includegraphics[width=1.0\linewidth]{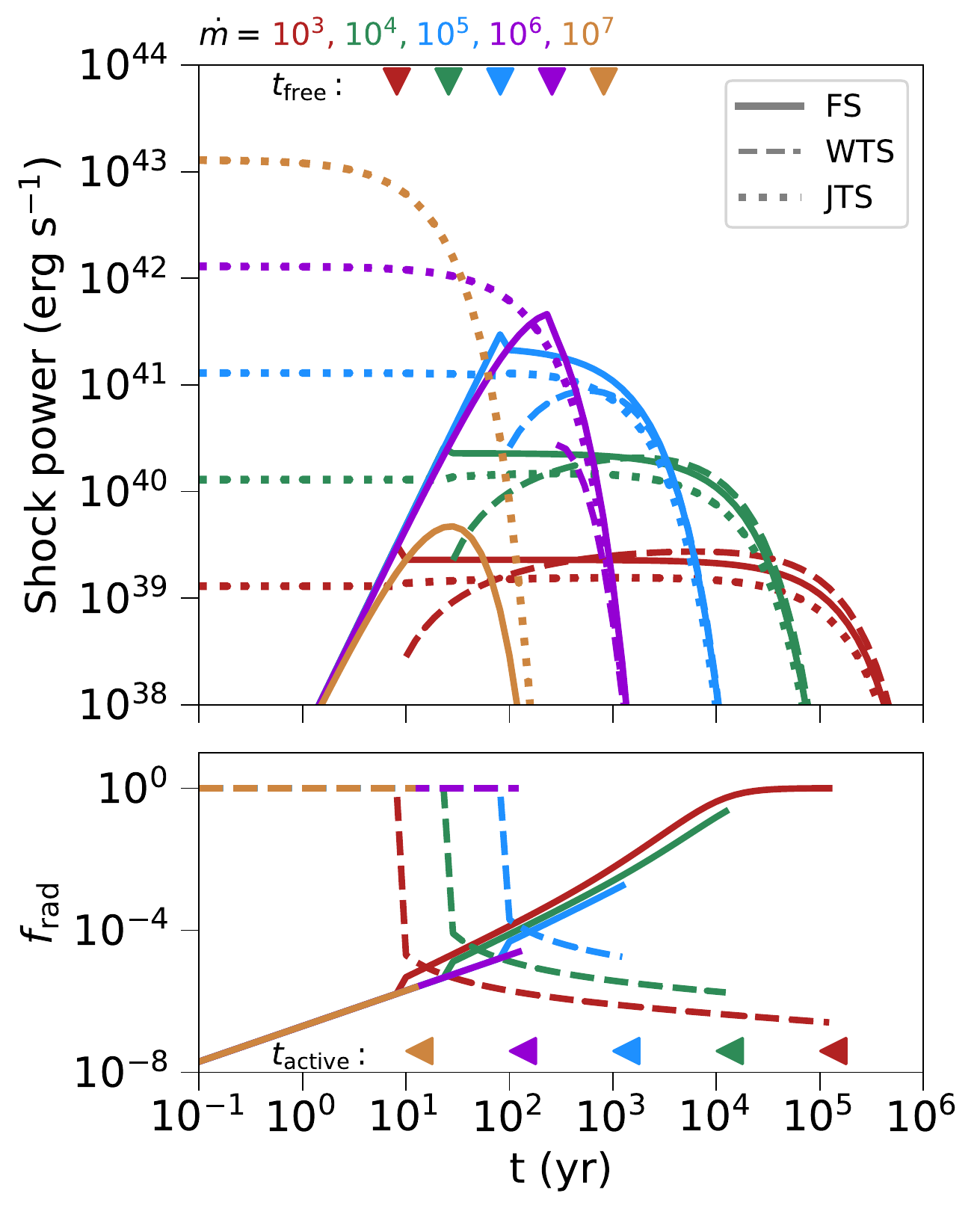}
\caption{Solid, dashed, and dotted curves show the properties of the forward shock, wind termination shock, and jet termination shock, respectively. The colors denote different accretion rates. The time evolution of the shock power (Eq.~\ref{eq:L_s}) is shown in the top panel, and the bottom panel shows the shocks' radiative efficiency (Eq.~\ref{eq:f_rad}). 
The left-facing triangles at the bottom frame denote the active lifetime of the accreting engine (Eq.~\ref{eq:t_active}), and the downward-facing triangles at the top frame denote the free-expansion timescale of the shock (Eq.~\ref{eq:t_free}).}
\label{fig:shock_power}
\end{figure}

The top panel of Fig.~\ref{fig:shock_power} shows the time evolution of the power in the three shocks, for different values of $\dot{m}$. For purposes of calculating the temporal evolution of various shock properties, the engine is smoothly `turned-off' at a time $t = t_{\rm active}$ when accretion stops (Eq.~\ref{eq:t_active}), by multiplying the shock properties with an exponential function of the form $e^{-t/t_{\rm active}}$.  At early times ($t\ll t_{\rm free}$), the power in the JTS dominates and remains nearly constant (as $v_{\rm j}\gg v_{\rm w} \gtrsim v_{\rm fs}$) until the engine shuts off.  Once the shell crosses the free-expansion phase at $t > t_{\rm free}$ (as occurs only for $\dot{m}<10^6$), the higher $\rho_{\rm fs}$ ($> \rho_{\rm wts}\gg\rho_{\rm jts}$) renders the kinetic power of the FS and WTS comparable to that of the JTS. The bottom panel of Fig.~\ref{fig:shock_power} shows the radiative efficiency ($f_{\rm rad}$) of the FS and WTS. (We do not show $f_{\rm rad}$ for the JTS because the cooling time of the relativistic particles is extremely long $>10^{10}$\,yr and hence $f_{\rm rad}\ll10^{-8}$.) The radiative efficiency of the FS and WTS generally increases with $\dot{m}$. However, unlike the FS, which turns from being adiabatic to radiative after a critical time $t=t_{\rm rad}^{\rm fs}$, the WTS turns from being radiative to adiabatic soon after its onset.  Also note that only for $\dot{m}\lesssim10^4$ does the FS become radiative during the accretion active time $t < t_{\rm active}$. 

For the rest of this paper, we shall focus on the JTS, for the following reason, motivated further in the sections to follow. Even though the total energy carried dissipated at the FS dominates that of either the JTS or the WTS, the JTS dominates the shock energy {\it along the narrow solid angle along the jet axis relevant for photohadronic interactions within the geometrically-beamed X-ray emission of the accretion flow} (see Eq.~\ref{eq:L_X} for further discussion).

The magnetic field strength in the vicinity of the JTS, can be estimated from the jet magnetization,
\begin{align}\label{eq:B_jts}
B_{\rm jts} & = \sqrt{16\pi\sigma_{\rm j}m_{\rm p}n_{\rm s} v_{\rm j}^2}\approx 4\,{\rm mG}\,\sigma_{\rm j,-1}^{1/2}\left(\frac{v_{\rm j}}{0.3\,c}\right)^{1/2}\eta_{-1}^{1/2} \nonumber \\
& \times 
    \begin{cases}
    \dot{m}_{5}^{1/2}\left(\frac{t}{\rm 80\,yr}\right)^{-1} & (t<t_{\rm free}) \\
    \dot{m}_{5}^{3/10}v_{\rm w,9}^{3/5}n_{1}^{1/5}\left(\frac{t}{\rm 80\,yr}\right)^{-3/5} & (t>t_{\rm free})
    \end{cases}
    ,
\end{align}
where $n_{\rm s}=4\rho_{\rm jts}/\mu m_{\rm p}$ is the post-JTS particle number density. 

\begin{figure} 
\centering
\includegraphics[width=1.0\linewidth]{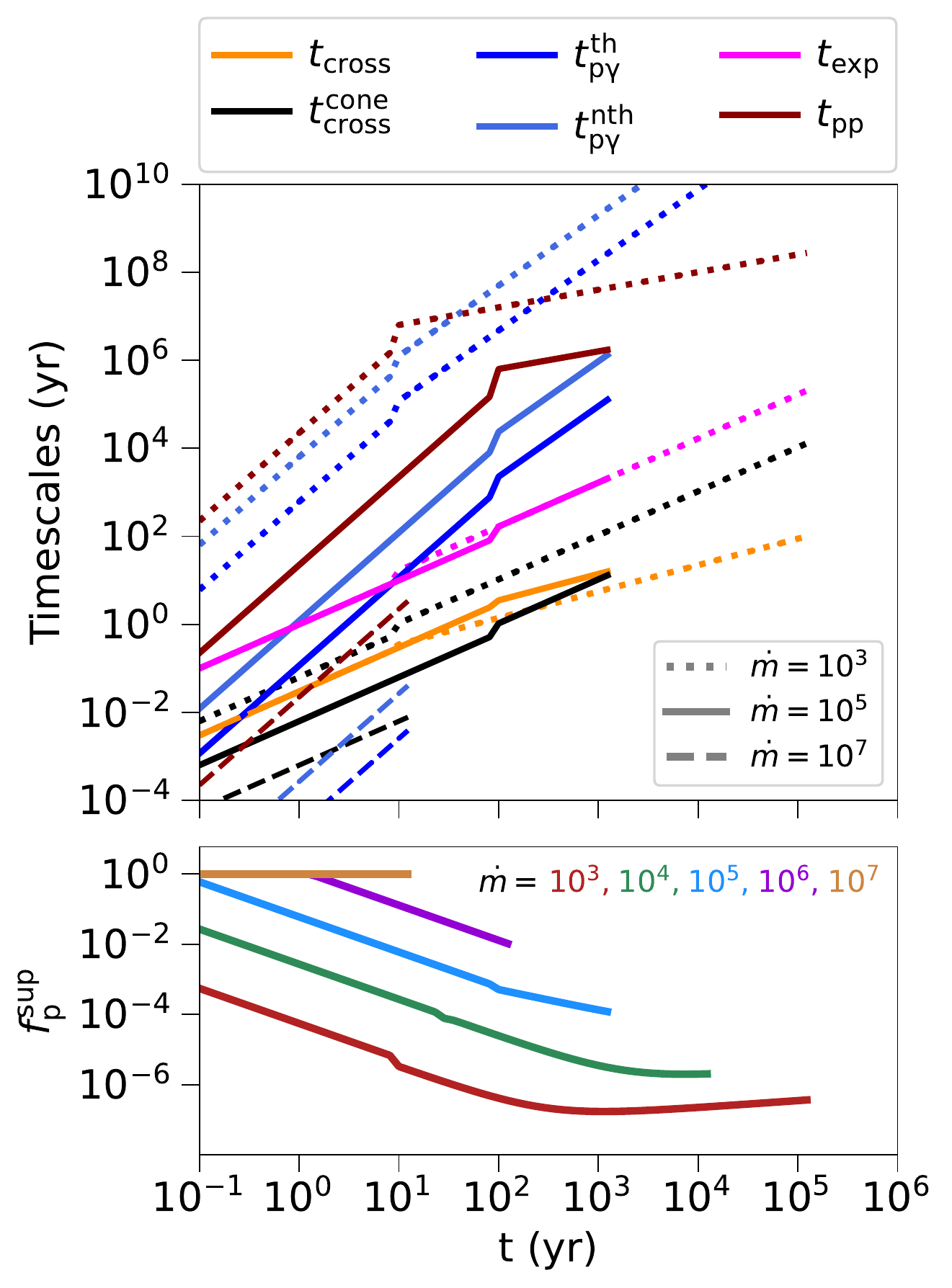}
\caption{\textit{Top panel:} Different timescales of the system for our fiducial model with $\dot{m}=10^5$ (solid curves); dotted and dashed curves represent $\dot{m}=10^3$ and $\dot{m}=10^7$, respectively. Pink curves denote the hypernebula expansion timescale (Eq.~\ref{eq:t_exp}); yellow curves denote the light-crossing timescale of the hypernebula ($t_{\rm cross}=R/c$), and the black curves denote the time particles take to laterally cross the X-ray emission cone (Eq.~\ref{eq:t_cross_cone}). Dark and light blue curves denote the interaction timescale of protons between thermal and nonthermal photons (Eq.~\ref{eq:t_pgamma}), respectively. Dark brown curves denote the hadronic ($pp$) interaction timescale (Eq.~\ref{eq:t_pp}). The overall pion creation timescale (Eq.~\ref{eq:t_pion}), not shown here, overlaps with $t_{\rm p\gamma}^{\rm th}$ at earlier times, and with $t_{\rm pp}$ at later times, when applicable (e.g., for $\dot{m}=10^3$).
\textit{Bottom panel:} Neutrino production suppression factor for different accretion rates for the fiducial model (Eq.~\ref{eq:f_psup}).}
\label{fig:timescales}
\end{figure}

\begin{figure} 
\centering
\includegraphics[width=1.0\linewidth]{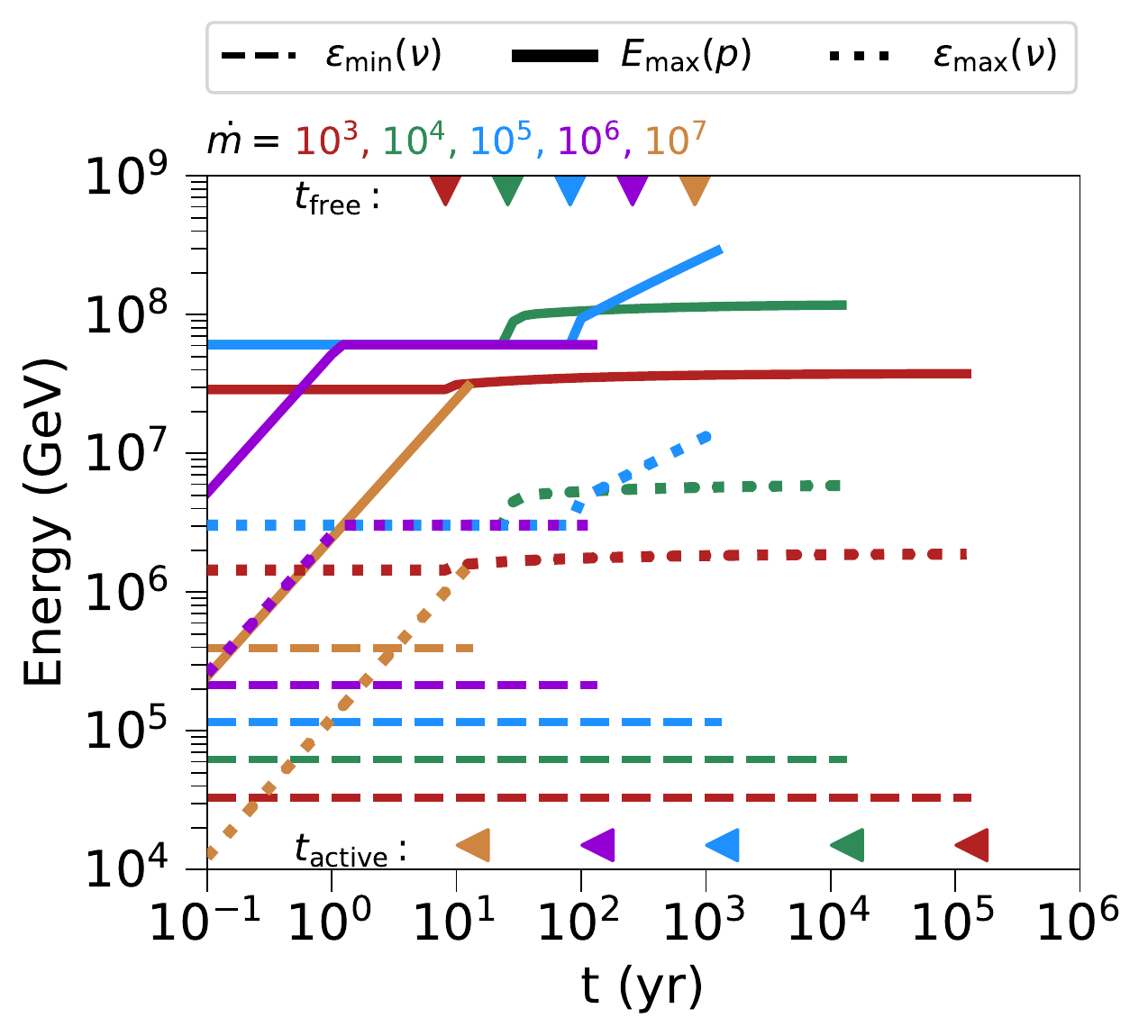}
\caption{Limitations on the proton and neutrino energies. Solid curves represent the maximum proton energy (Eq.~\ref{eq:E_max}); the corresponding maximum attainable neutrino energy is denoted by dotted curves. Dashed curves denote the minimum attainable neutrino energy through the photomeson process (Eq.~\ref{eq:varepsilon_min}). Different colors represent different accretion rates. The left-facing triangles in the lower frame represent the active duration of the engine ($t_{\rm active}$; Eq.~\ref{eq:t_active}) for different $\dot{m}$, and the bottom-facing triangles along the upper frame mark the end of the free expansion of the shell ($t_{\rm free}$; Eq.~\ref{eq:t_free}).}
\label{fig:energies}
\end{figure}

\subsection{Background radiation} \label{subsec:background_radiation}

There are at least three sources of background photons\footnote{We do not consider thermal synchrotron radiation that peaks in the radio band \citep{Sridhar&Metzger22} because, the required proton threshold energy of $E_{\rm thr}\sim10$\,ZeV to interact with $\sim$GHz photons and produce pions (Eq.~\ref{eq:E_thr}) is much larger than the maximum energy to which protons can be accelerated at the hypernebula JTS (Fig.~\ref{fig:energies}).} with which relativistic ions accelerated at the JTS (\S\ref{subsec:proton_energization}) can interact to produce neutrinos.

\textit{(1) Disk thermal emission:} 
For accretion rates $\dot{m}\gg1$, advection and outflows from the accretion disk limit the bolometric luminosity of the blackbody emission to  \citep{Begelman+06, Poutanen+07},
\be \label{eq:L_bol}
L_{\rm bol} = L_{\rm Edd}(1+\ln\dot{m}) \approx 10^{40}\,{\rm erg\,s}^{-1}(1 + \ln \dot{m}_{5}).
\ee
The isotropic-equivalent luminosity of the disk emission along the axis of the inner accretion funnel, however, can greatly exceed this value due to geometric collimation by the accretion flow (e.g., \citealt{King09}). For an X-ray emitting cone of half-opening angle $\theta_{\rm X} \ll 1$, the corresponding beaming fraction can be defined as,
\be \label{eq:f_b}
f_{\rm b,X} \equiv 2\pi[1 - \cos{(\theta_{\rm X})}]/4\pi \simeq \theta_{\rm X}^2/4.
\ee
Motivated by radiation magnetohydrodynamical simulations of super-Eddington accretion (\citealt{Sadowski&Narayan15}, who found $f_{\rm b,X} \approx 10^{-3}$ for $\dot{m} \sim 10^{3}$), we conservatively adopt a beaming fraction $f_{\rm b,X}\approx 10^{-3}(\dot{m}/10^{3})^{-1}$, shallower than the $f_{\rm b,X} \propto \dot{m}^{-2}$ dependence advocated by \citet{King09}.
This gives an isotropic-equivalent X-ray luminosity (within angles $\theta < \theta_{\rm X})$, 
\be \label{eq:L_X}
L_{\rm X}=\frac{L_{\rm bol}}{f_{\rm b,X}}\approx10^{45}\,{\rm erg\,s}^{-1}(1 + \ln \dot{m}_{5})\dot{m}_{5}.
\ee
The high densities of the jetted outflows permeating the collimated accreting funnel further imply that thermal photons emerge not directly from the disk surface, but rather from the fast wind/jet photosphere at much larger radii.  Assuming the kinetic luminosity of the fast wind to be comparable to its photon luminosity \citep{Sadowski&Narayan15}, i.e. $\dot{M}_{\rm j}v_{\rm j}^{2}/2 \approx L_{\rm X}$, the photosphere radius can be approximated as
\be \label{eq:r_ph}
r_{\rm ph} = \frac{L_{\rm X}\sigma_{\rm T}(1-v_{\rm j}/c)}{4\pi m_{\rm p}c^2v_{\rm j}} \approx 10^{12}\,{\rm cm}\,(1 + \ln \dot{m}_{5})\dot{m}_{5} \left(\frac{v_{\rm j}}{0.3\,c}\right)^{-1},
\ee
where $\sigma_{\rm T}$ is the Thomson cross section.  The resulting effective temperature of the thermal emission is given by,
\begin{align} \label{eq:kBTeff}
k_{\rm B}T_{\rm eff} & = \left(\frac{L_{\rm X}}{4\pi r_{\rm ph}^2\sigma_{\rm SB}}\right)^{1/4} \approx 60\,{\rm eV} \nonumber\\ 
&\times(1 + \ln \dot{m}_{5})^{-1/4}\dot{m}_{5}^{-1/4} \left(\frac{v_{\rm j}}{0.3\,c}\right)^{1/2},
\end{align}
where, $\sigma_{\rm SB}=5.67\times10^{-5}\,{\rm g\,s^{-3}\,K^{-4}}$ is the Stefan-Boltzmann constant. 
Interestingly, hypernebulae could thus manifest as ultraluminous supersoft X-ray sources \citep{Kahabka_06, Micic+22}.

The number density of the thermal disk photons at the radius of the jet termination shock is thus given by,
\begin{align} \label{eq:n_X_th}
    n_{\rm X}^{\rm th} & \approx \frac{L_{\rm X}}{4\pi R^2k_{\rm B}T_{\rm eff}c} = 4\times10^6\,{\rm cm}^{-3}\,(1 + \ln \dot{m}_{5})^{5/4}\dot{m}_{5}^{5/4} \nonumber \\
    & \times \left(\frac{v_{\rm j}}{0.3\,c}\right)^{-1/2} \times
    \begin{cases}
        v_{\rm w,9}^{-2}\left(\frac{t}{80\,{\rm yr}}\right)^{-2} & ({\rm t < t_{\rm free}})\\
        \left(\frac{L_{\rm w,42}}{n_{1}}\right)^{-2/5}\left(\frac{t}{80\,{\rm yr}}\right)^{-6/5} & (t > t_{\rm free})
    \end{cases}
    .
\end{align}

\textit{(2) Disk/Comptonized nonthermal emission:} The thermal UV/X-ray disk emission may be up-scattered by a Comptonizing corona. Given the large Thomson depth in the polar funnel, the Comptonized nonthermal emission would be generated in the funnel/jet walls at large distances $>r_{\rm ph}$ (e.g., powered by processes such as magnetic reconnection \citealt{Sironi&Beloborodov20, sridhar+21c, Sridhar+23a}) along the jet boundary \citep{Ripperda+20}. The Comptonized photons follow a powerlaw spectrum $N(\epsilon)\propto\epsilon^{-q}$ with say, a powerlaw index $q=2$, from the thermal peak energy $\epsilon_{\rm pk,th}^{\rm eff}=3k_{\rm B}T_{\rm eff}$ to the pair creation threshold energy $2m_{\rm e}c^2$ \citep{Svensson87}. Assuming that the energy contained in the nonthermal photons is comparable to the thermal photons, the number density of the nonthermal photons can be estimated as,
\begin{align}  \label{eq:n_X_nonth}
n_{\rm X}^{\rm nth} & \sim\frac{L_{\rm X}}{4\pi R^2 \epsilon_{\rm pk,th}^{\rm eff}c\ln{(2m_{\rm e}c^2/\epsilon_{\rm pk,th}^{\rm eff})}} \nonumber \\
& \approx 4\times10^5\,{\rm cm}^{-3}\,\left(\frac{v_{\rm j}}{0.3\,c}\right)^{-1/2} (1 + \ln \dot{m}_{5})^{5/4}\dot{m}_{5}^{5/4}\nonumber \\
    & \times 
    \begin{cases}
        v_{\rm w,9}^{-2}\left(\frac{t}{80\,{\rm yr}}\right)^{-2} & ({\rm t < t_{\rm free}})\\
        \left(\frac{L_{\rm w,42}}{n_{1}}\right)^{-2/5}\left(\frac{t}{80\,{\rm yr}}\right)^{-6/5} & (t > t_{\rm free})
    \end{cases}
    .
\end{align}

\textit{(3) Free-free X-ray emission:} 
The FS becomes radiative after a time $t>t_{\rm rad}^{\rm fs}$ when $f_{\rm rad} \simeq 1$ (see bottom panel of Fig.~\ref{fig:shock_power}), and its luminosity dominates that of both the WTS and JTS, which remain adiabatic.  The number density of photons near the FS can be estimated as
\be  \label{eq:n_X_free}
    n_{\rm X}^{\rm ff} = \frac{\Lambda_{\rm ff}}{\Lambda}\frac{f_{\rm rad}L_{\rm s}}{4\pi R^2\bar{E}c},
\ee
where the factor $\Lambda_{\rm ff}/\Lambda$ accounts for the fraction of the FS's total luminosity radiated via free-free emission, which typically dominates over line cooling given the high post-shock temperature $T>10^7$\,K (Eq.~\ref{eq:mean_temp}).

\begin{figure} 
\centering
\includegraphics[width=1.0\linewidth]{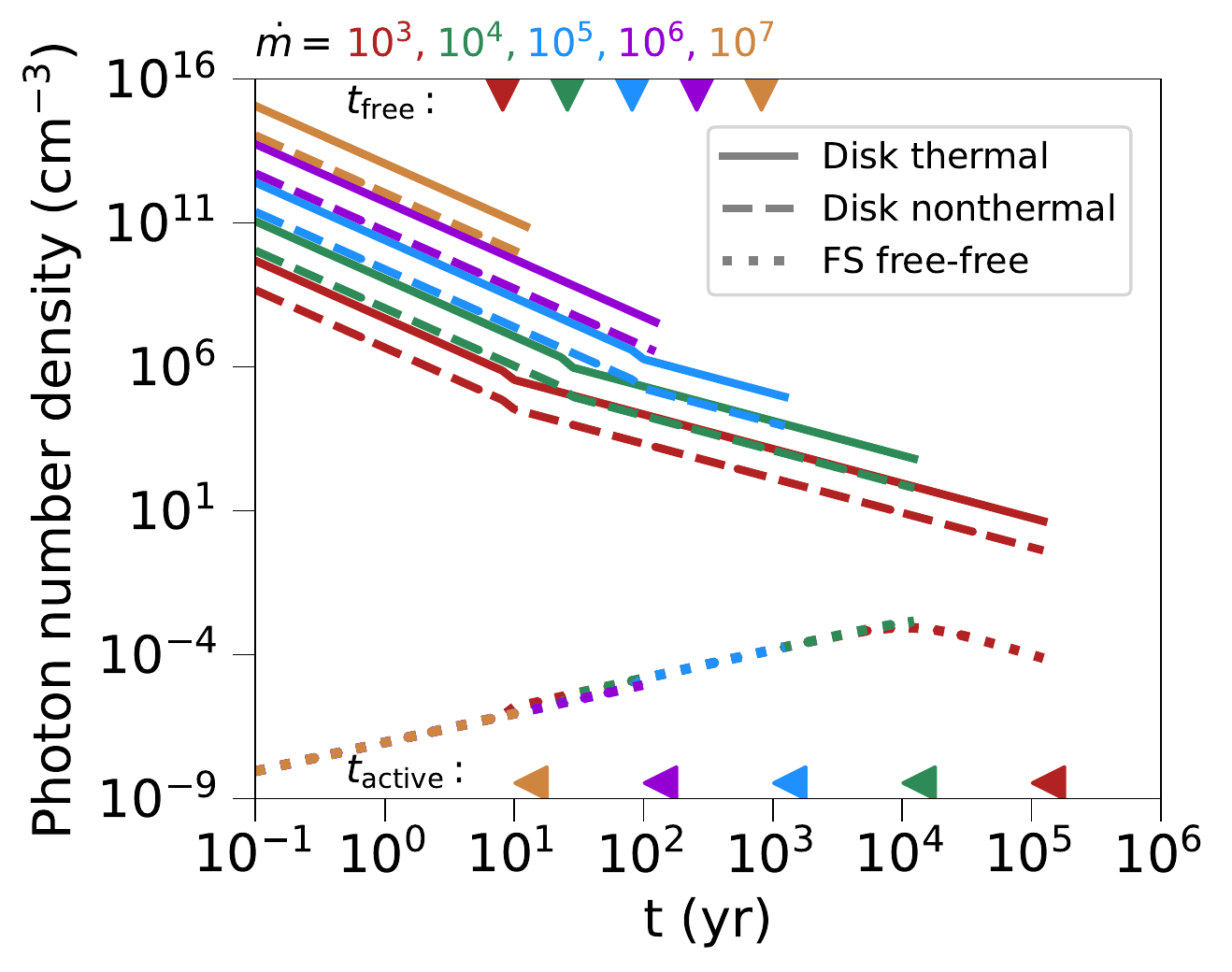}
\caption{Background UV/X-ray photon number density near the location of the JTS. Solid curves represent the disk blackbody photons (Eq.~\ref{eq:n_X_th}), dashed curves represent the Compton upscattered nonthermal photons (Eq.~\ref{eq:n_X_nonth}), and the dotted curves represent the free-free X-ray photons (Eq.~\ref{eq:n_X_free}). Different colors indicate different accretion rates. The left-facing triangles in the lower frame represent the active duration of the engine ($t_{\rm active}$; Eq.~\ref{eq:t_active}) for different $\dot{m}$, and the bottom-facing triangles along the upper frame mark the end of the free expansion of the shell ($t_{\rm free}$; Eq.~\ref{eq:t_free}).}
\label{fig:background_radiation}
\end{figure}

The photon number densities from the aforementioned three emission processes are shown in Fig.~\ref{fig:background_radiation}. Although $n_{\rm X}^{\rm ff}$ increases with time as $\propto t^{4/5}$ whereas $n_{\rm X}^{\rm th}$ and $n_{\rm X}^{\rm nth}$ decreases as $\propto t^{-6/5}$, we find $n_{\rm X}^{\rm th} > n_{\rm X}^{\rm nth} \gg n_{\rm X}^{\rm ff}$ at all times $t<t_{\rm active}$ for all $\dot{m}$.  Thus motivated, in what follows we shall focus on the interaction of ions accelerated at the JTS with the thermal disk emission and its nonthermal Comptonized counterpart.  The densities of the other sources of hard X-ray/gamma-ray photons e.g., from the jet knots \citep{Safi-Harb+22} are also negligible compared to $n_{\rm X}^{\rm th/nth}$.

\section{Model for neutrino emission} \label{sec:neutrino_model}

\subsection{Pion production} \label{subsec:pion_production}

Immersed in an UV/X-ray photon field (\S\ref{subsec:background_radiation}), the protons accelerated at the JTS can engage in the photomeson process through the $\Delta^+$-resonance \citep{Berezinskii+90}, 
\begin{eqnarray} \label{eq:pgamma}
	p+\gamma  & \longrightarrow &  \Delta^+ \  \longrightarrow \  \left\{
		\begin{array}{l l}
			p + \pi^0 &  \\ 
			n + \pi^+ &  \\ 
		\end{array} 
	\right. \ . 
\end{eqnarray} 
The $\Delta^+$-resonance decay branching ratio is such that two-thirds of the products follow the neutral pion ($\pi^0$) channel, and the remaining one-thirds will follow the charged pion ($\pi^+$) channel. The pion production through the photohadronic process (Eq.~\ref{eq:pgamma}) with thermal photons from the accretion disk peaks at the $\Delta^+$-resonance at a threshold proton energy given by \citep{Mannheim&Biermann89, Waxman&Bahcall97},
\be \label{eq:E_thr}
E_{\rm thr} \gtrsim \frac{E_{\Delta^+}}{k_{\rm B}T_{\rm eff}}\frac{m_{\rm p}c^2}{2}\approx 2\,{\rm PeV}(1 + \ln \dot{m}_{5})^{1/4}\dot{m}_{5}^{1/4} \left(\frac{v_{\rm j}}{0.3\,c}\right)^{-1/2},
\ee
where $E_{\Delta^+}\approx0.3$\,GeV is the energy where the cross-section peaks. However, note that $E_{\rm thr}$ can be smaller by a factor $(\epsilon/2m_{\rm e}c^2)^{-2}$ for photohadronic interactions with higher-energy nonthermal photons with energies $k_{\rm B}T_{\rm eff}{<}\epsilon{<}2m_{\rm e}c^2$ (more discussion around Eq.~\ref{eq:varepsilon_min}). Since $E_{\rm thr}\gg \bar{E}$, the photohadronic reactions with thermal photons can proceed only with the JTS-accelerated protons from the nonthermal tail with energies $E\gg\bar{E}$ (more in \S\ref{subsec:proton_energization}). 

In addition to this channel, pions will also be produced through $pp$-interactions that follow,
\begin{eqnarray} \label{eq:pp}
	p+p  & \longrightarrow &   \left\{
		\begin{array}{l l}
			  \pi^+ + X \\ 
			 \pi^0 + X\\
                 \pi^-  + X\\
		\end{array}  
	\right. \ , 
\end{eqnarray} 
where $X$ refers to any other by-products (such as $\phi$-mesons, K-mesons, pions, protons, and neutrons) produced in the reaction other than the particle indicated. In the following section, we calculate the particle interaction timescales that will determine the efficiency of pion production through $p\gamma$ (Eq.~\ref{eq:pgamma}) and $pp$ channels (Eq.~\ref{eq:pp}).  

\subsection{Particle interaction timescales} \label{subsec:interaction_timescales}

The most energetic particles will propagate inside the magenetized region of the hypernebula on the light-crossing time, $t_{\rm cross}=R/c$, while most of the less-energetic particles that are trapped inside the nebula by the turbulent magnetic field would escape the system on a longer, shell expansion timescale (Eq.~\ref{eq:t_exp}). The particles that are the most relevant for neutrino production are the ones that are trapped within the X-ray solid angle (Eq.~\ref{eq:f_b}) and can engage in photohadronic interactions. We can correspondingly define a timescale for laterally crossing the X-ray cone as,
\be \label{eq:t_cross_cone}
t_{\rm cross}^{\rm cone} \simeq \frac{R}{v_{\rm fs}}\theta_{\rm X} \approx 0.5\,{\rm yr}\left(\frac{t}{\rm 80\,yr}\right)\dot{m}_{5}^{-1}.
\ee

Let's first consider the photohadronic interactions of the protons accelerated at the JTS with the thermal and nonthermal photons (Eqs.~\ref{eq:n_X_th} and \ref{eq:n_X_nonth}). We can define an optical depth for this interaction, $\tau_{\rm p\gamma}^{\rm (th,nth)} = \sigma_{\rm p\gamma}\kappa_{\rm p\gamma}Rn_{\rm X}^{\rm (th,nth)}$ where, $\sigma_{\rm p\gamma}\sim5\times10^{-28}\,{\rm cm}^2$ is the cross-section of the photopion production \citep{ParticleDataGroup:2004fcd} at the $\Delta^+$-resonance, and $\kappa_{\rm p\gamma}\sim0.15$ is the average fraction of energy lost from a proton per elastic collision.  This can be deemed the photopion production efficiency: i.e., the fraction of the shock-accelerated protons that will be converted into pions before streaming out of the shock, so long as they satisfy the threshold energy requirements (Eq.~\ref{eq:E_thr}). 

The corresponding $p\gamma$ interaction timescale is,
\begin{align} \label{eq:t_pgamma}
& t_{\rm p\gamma}^{\rm (th,nth)} = \frac{t_{\rm cross}^{\rm cone}}{\tau_{\rm p\gamma}^{\rm (th,nth)}} \approx 8(1 + \ln \dot{m}_{5})^{-5/4}\dot{m}_{5}^{-7/4}\left(\frac{v_{\rm j}}{0.3\,c}\right)^{1/2}\nonumber \\
    & \times
    \begin{cases}
        (10^2,10^3)\,{\rm yr}\,v_{\rm w,9}\left(\frac{t}{80\,{\rm yr}}\right)^{2} & ({\rm t < t_{\rm free}})\\
        (10^2,10^3)\,{\rm yr}\,\left(\frac{L_{\rm w,42}}{n_{1}}\right)^{1/5}\left(\frac{t}{80\,{\rm yr}}\right)^{8/5} & (t > t_{\rm free})
    \end{cases}
    .
\end{align}
The other relevant channel of neutrino production is the hadronuclear ($pp$) interaction of protons accelerated by the JTS, whose interaction timescale is,
\begin{align} \label{eq:t_pp}
t_{\rm pp} & \simeq \frac{\theta_{\rm X}}{n_{\rm p}\sigma_{\rm pp}\kappa_{\rm pp}v_{\rm fs}} 
\approx 1.5\times10^{5}\,{\rm yr} \nonumber \\
&\times
    \begin{cases}
          \dot{m}_{5}^{-3/2}v_{\rm w}^{2}\left(\frac{t}{80\,{\rm yr}}\right)^{2} & (t < t_{\rm free})\\
          \dot{m}_{5}^{-1/2}\left(\frac{L_{\rm w,42}}{n_{1}}\right)^{-1/5}n_1^{-1}\left(\frac{t}{80\,{\rm yr}}\right)^{2/5} & (t > t_{\rm free})
    \end{cases}
    ,
\end{align}
where $\sigma_{\rm pp}\sim10^{-25}\,{\rm cm}^{-2}$ (at $\sim$1\,EeV) and $\kappa_{\rm pp}\sim0.5$ \citep{ParticleDataGroup:2004fcd}. Note that $t_{\rm pp}$ increases sharply as $\propto t^{2}$ only until $t=t_{\rm free}$, beyond which it increases moderately as $\propto t^{2/5}$, given now the constant influx of protons into the shell from the CSM (see Eq.~\ref{eq:n_p}). The total pion creation rate due to photopion and hadronuclear interactions can be written as,
\be \label{eq:t_pion}
1/t_{\pi,\rm cre} = 1/t_{\rm p\gamma}^{\rm th} + 1/t_{\rm p\gamma}^{\rm nth} + 1/t_{\rm pp}.
\ee 

The top panel of Fig.~\ref{fig:timescales} shows the aforementioned interaction timescales for different accretion rates ($\dot{m}=10^3$: dotted curves, $\dot{m}=10^5$: solid curves, and $\dot{m}=10^7$: dashed curves). At all times, for all choices of $\dot{m}$, $t_{\rm cross} < t_{\rm exp}$. However, $t_{\rm cross}^{\rm cone}$ is dependent on $\dot{m}$ through the beaming fraction (Eq.~\ref{eq:f_b}): $t_{\rm cross}^{\rm cone} < t_{\rm cross}$ for $\dot{m}\gtrsim10^4$. For low $\dot{m}\sim10^3$ systems, the timescales follow the hierarchy, $t_{\rm exp} \ll t_{\rm p\gamma}^{\rm th} < t_{\rm p\gamma}^{\rm nth} < t_{\rm pp}$ at earlier times $t \lesssim t_{\rm free}$, and at later times $t\gg t_{\rm free}$, the constant shell density changes the hierarchy to $t_{\rm pp} < t_{\rm p\gamma}^{\rm th} < t_{\rm p\gamma}^{\rm nth}$. For higher $\dot{m}\sim10^7$ cases, the timescale hierarchy obeys $t_{\rm p\gamma}^{\rm th} < t_{\rm p\gamma}^{\rm nth} < t_{\rm cross} < t_{\rm pp} < t_{\rm exp}$ during most of their lifetimes.  Overall, (although not shown in Fig.~\ref{fig:timescales}), $t_{\rm \pi,cre}$ overlaps with $t_{\rm p\gamma}^{\rm th}$ at earlier times and with $t_{\rm pp}$ at later times (if applicable, e.g., for low $\dot{m}\sim10^3$ cases). 

Pion production happens efficiently as long as $t_{\pi,\rm cre}$ is smaller than both the $t_{\rm exp}$ and the synchrotron cooling timescale of the proton given by,
\begin{align} \label{eq:t_psyn}
t_{\rm p,syn} &= \frac{6m_{\rm p}^3c}{\sigma_{\rm T}m_{\rm e}^2\bar{\gamma}_{\rm p}B_{\rm sh}^2}  \sim 10^{10}\,{\rm yr}\, \sigma_{\rm j,-1}^{-1}\left(\frac{v_{\rm j}}{0.3\,c}\right)^{-1}\eta_{-1}^{-1} \nonumber \\
& \times 
    \begin{cases}
    \dot{m}_{5}^{-1}\left(\frac{t}{\rm 80\,yr}\right)^{2} & (t<t_{\rm free}) \\
    \dot{m}_{5}^{-3/5}v_{\rm w,9}^{-6/5}n_{1}^{-2/5}\left(\frac{t}{\rm 80\,yr}\right)^{6/5} & (t>t_{\rm free})
    \end{cases}
\end{align}
where we take $\bar{\gamma}_{\rm p}\sim10^6$ to be the average Lorentz factor of the accelerated protons with energy in the relevant energy range for neutrino production ($E_{\rm thr} < E < E_{\rm max}$; more in Eqs.~\ref{eq:E_max} and \ref{eq:E_thr}). Comparing the curves in the top panel of Fig.~\ref{fig:timescales} demonstrates the generally weak impact $t_{\rm p,syn}$ has on any of the relevant timescales for pion production.
Given these timescales, the overall efficiency of pion production can be quantified by defining a `suppression factor',
\be \label{eq:f_psup}
f^{\rm p}_{\rm sup} = {\rm min}\left[1,\frac{t_{\rm cross}}{t_{\pi,\rm cre}}, \frac{t_{\rm cross}^{\rm cone}}{t_{\pi,\rm cre}}, \frac{t_{\rm p,syn}}{t_{\pi,\rm cre}}\right].
\ee
As the bottom panel of Fig.~\ref{fig:timescales} shows, the neutrino production is not suppressed (i.e., $f_{\rm p}^{\rm sup}\sim1$) for high $\dot{m}\gtrsim10^6$ systems throughout their lifetimes. This is because of the large background photon density due to a stronger beaming effect that reduces the photohadronic interaction timescales. On the other hand, for weaker accretion rates $\dot{m}\lesssim10^6$ the pion production efficiency decreases with time as $f_{\rm sup}^{\rm p}\propto t^{-1}$ until $t=t_{\rm free}$ owing to its expansion and smaller background photon density. Soon after, the shell reaches a constant density (leading to a modestly increasing $t_{\rm pp}$), and the consequently decreasing $t_{\rm pp}/t_{\rm exp}$ implies that $t_{\rm \pi,cre}$ becomes dominated by hadronic interactions. This effect is most pronounced for the long-lived $\dot{m}=10^3$ case where $f_{\rm p}^{\rm sup}$ increases again at later times. Overall, $f_{\rm p}^{\rm sup}\ll1$ for the majority of the lifetimes of low $\dot{m}\lesssim10^6$ systems.       

An added complication is that pions and muons could suffer synchrotron losses (pion/muon `damping') which inhibit neutrino production if the former proceeds at a timescale shorter than the latter.  We assess this possibility by calculating the critical energy above which their synchrotron losses dominate \citep{Waxman&Bahcall98, Metzger+20},
\begin{align} \label{eq:Esyn_pimu}
& \varepsilon_{\pi,\mu}^{\rm syn} = \left(\frac{6\pi m_{\pi,\mu}c}{\sigma_{\rm T}B_{\rm sh}^2\tau_{\pi,\mu}}\right)^{1/2}\left(\frac{m_{\pi,\mu}}{m_{\rm e}}\right)m_{\pi,\mu}c^2 \nonumber \\
& \approx (30,2)\,{\rm ZeV}\,\sigma_{\rm j,-1}^{-1/2} \times \left(\frac{v_{\rm j}}{0.3\,c}\right)^{-1/2}\eta_{-1}^{-1/2} \nonumber \\
&  \times
    \begin{cases}
    \dot{m}_{5}^{-1/2}\left(\frac{t}{\rm 80\,yr}\right) & (t<t_{\rm free}) \\
    \dot{m}_{5}^{-3/10}v_{\rm w,9}^{-3/5}n_{1}^{-1/5}\left(\frac{t}{\rm 80\,yr}\right)^{3/5} & (t>t_{\rm free})
    \end{cases}
    ,
\end{align}
where $m_\pi$ and $m_\mu$ are the masses of pions and muons, and $\tau_\pi=2.6\times10^{-8}$\,s and $\tau_\mu=2.2\times10^{-6}$\,s are their decay times \citep{ParticleDataGroup:2004fcd}. These energies---as we show below in Eq.~\ref{eq:E_max}---exceed the maximum attainable proton energies during the life of any hypernebula. Therefore, it is fair to assume that majority of the neutrino production from the decay of pions and muons completes before their radiative losses take over.  Furthermore, due to the small proton number density at the shock radius (see Eq.~\ref{eq:n_p}), the timescales of pion and muon hadronuclear interactions, $t_{\rm \pi,p}$ and $t_{\rm \mu,p}$ greatly exceed their lifetimes $\tau_{\pi}$ and $\tau_{\mu}$. Therefore, the efficiency of neutrino production is not additionally suppressed.

\subsection{Proton energization}\label{subsec:proton_energization}

A fraction of the luminosity of the JTS is used to heat protons (to an average energy $\bar{E}\simeq17$\,MeV; Eq.~\ref{eq:mean_temp}) and accelerate them into a power-law distribution of relativistic energies $E\gg\bar{E}$ via the diffusive shock acceleration mechanism \citep[DSA;][]{Blandford&Ostriker78},
\be \label{eq:f_E}
    f_{\rm E}=
    \begin{cases}
           1/2 & (E\sim \bar{E}) \\
           \epsilon_{\rm rel} & (E\gg \bar{E}),
    \end{cases}
\ee
where $\epsilon_{\rm rel}\lesssim0.01$ is the efficiency of the non-thermal acceleration process \citep[e.g., for mildly-relativistic shocks;][]{Fang+20}.\footnote{Note that nonthermal particle acceleration can be inefficient in highly magnetized shocks \citep{Sironi&Spitkovsky11}.} The time evolution of the proton luminosity, $L_{\rm p}(E)=f_{\rm E}(E)L_{\rm s}$ for different $\dot{m}$ is shown in  Fig.~\ref{fig:luminosities}.

The accelerated distribution of non-thermal protons can be described by,
\be \label{eq:particle_powerlaw}
\frac{dN_{\rm p}}{dE} \sim \left(\frac{E}{\bar{E}}\right)^{-q}e^{-E/E_{\rm max}}.
\ee
where we take $q\simeq2$ \citep[our fiducial case;][]{Blandford&Eichler87}; steeper spectra with $q\gtrsim2$ could also arise as the magnetic fluctuations behind the FS drift away at the local Alfvén speed \citep[`postcursor';][]{Diesing+21}. 

The maximum energy $E_{\rm max}$ attained by the protons in the DSA process is determined by the competition between the diffusive acceleration timescale of particles with Larmor radius $r_{\rm L}$ \citep[e.g.,][]{Caprioli&Spitkovsky14},
\be \label{eq:t_acc}
t_{\rm acc}\simeq \frac{2}{3}\left(\frac{r_{\rm L}}{c}\right)\left(\frac{v_{\rm j}-v_{\rm fs}}{c}\right)^{-2} = \frac{2Ec}{3eB_{\rm sh}(v_{\rm j}-v_{\rm fs})^2},
\ee
and the timescales corresponding to various interaction and cooling processes (e.g., in Fig.~\ref{fig:timescales}). This yields a maximum ion energy of,
\begin{equation} \label{eq:E_max}
E_{\rm max} = \frac{3eB_{\rm sh}(v_{\rm j}-v_{\rm fs})^2}{2c}{\rm min}\left[t_{\rm cross},t_{\rm cross}^{\rm cone}, t_{\rm exp}, t_{\rm p\gamma}^{\rm (th,nth)}, t_{\rm pp}\right].
\end{equation}
Eq.~\ref{eq:E_max} provides a system-size-limited maximum energy of the particles considering that the energized particles with $r_{\rm L}>R$ (and the X-ray cone lateral radius) would escape the accelerator \citep{Hillas_84} and do not contribute to the neutrino production.
Solid curves in Fig.~\ref{fig:energies} show the evolution of $E_{\rm max}(t)$ for different $\dot{m}$---complicated by the influence of the X-ray beaming cone's opening angle on various particle interaction timescales. 
The highest energy attained in $\dot{m}=10^3$ and $10^7$ cases is $\sim30$\,PeV, and the highest energy attained overall is, $\sim300$\,PeV, in the $\dot{m}=10^5$ case. 

Fig.~\ref{fig:protonspectra} shows example ($q=2$) snapshots of the proton spectra for different $\dot{m}$, taken at the end of the free-expansion and active phases (times $t_{\rm free}$ and $t_{\rm active}$, respectively).

\subsection{Pion decay to neutrinos} \label{subsec:neutrino_production}

The pions produced through the $p\gamma$ (Eq.~\ref{eq:pgamma}) and $pp$ (Eq.~\ref{eq:pp}) channels will eventually decay into neutrinos and gamma-rays following,
\begin{subequations} \label{eq:pion_decay}
\begin{gather}
    \begin{align}
        \pi^0 \longrightarrow & \gamma\gamma \\
        \pi^+ \longrightarrow & \mu^+ + \nu_\mu \\
        & \rotatebox[origin=c]{180}{$\Lsh$} e^+ + \nu_{\rm e} + \bar{\nu}_\mu \\
        \pi^- \longrightarrow & \mu^- + \bar{\nu}_\mu \\
        & \rotatebox[origin=c]{180}{$\Lsh$} e^- + \bar{\nu}_{\rm e} + \nu_\mu
    \end{align}
\end{gather}
\end{subequations}
In the absence of significant pion/muon damping (Eq.~\ref{eq:Esyn_pimu}), the ratio of the neutrino flavors resulting from pion decay, post-oscillation, is $\nu_{\rm e} : \nu_{\mu} : \nu_{\tau} = 1 : 1 : 1$ by the time they reach Earth \citep{Becker08}.

Let's now consider the photomesonic process with thermal background photons, where the produced pions decay into neutrinos with $\sim5\%$ of the proton's energy $\varepsilon\approx E/20$ \citep{Stecker68, Kelner&Aharonian08}. This sets the lowest attainable neutrino energy to be,
\begin{align} \label{eq:varepsilon_min}
\varepsilon_{\rm min} & \approx {\rm max}[ \bar{E}/20, E_{\rm thr}/20] \nonumber \\
& \approx 100\,{\rm TeV}(1 + \ln \dot{m}_{5})^{1/4}\dot{m}_{5}^{1/4} \left(\frac{v_{\rm j}}{0.3\,c}\right)^{-1/2}.
\end{align}
Note that $\varepsilon_{\rm min}$ can be smaller by a factor $(\epsilon/2m_{\rm e}c^2)^{-2}$ for photohadronic interactions with nonthermal photons with energies $k_{\rm B}T_{\rm eff}{<}\epsilon{<}2m_{\rm e}c^2$. This would effectively place the minimum neutrino energy cut-off at $\varepsilon_{\rm min}[\epsilon_{\rm pk,th}^{\rm eff}/k_{\rm B}T_{\rm eff}]^{-2}\sim10$\,TeV (see Eq.~\ref{eq:E_thr} and also \S\ref{subsec:background_radiation}(2)).  Production of even lower energy neutrinos mediated by interactions with nonthermal photons of energies $\epsilon>\epsilon_{\rm pk,th}^{\rm eff}$ is highly suppressed. On the other hand, the absolute maximum neutrino energy as set by the accelerator is $\varepsilon_{\rm max}\sim E_{\rm max}/20$.
Fig.~\ref{fig:energies} shows $\varepsilon_{\rm min}$ as the horizontal dashed line, which is independent of the age of the hypernebula, and cares only about $k_{\rm B}T_{\rm eff}$ through $\dot{m}$ (Eq.~\ref{eq:kBTeff}). $\varepsilon_{\rm max}$ is denoted by dotted curves, which trace $E_{\rm max}$ with a scaling offset of 1/20. In general, hypernebulae generate neutrinos throughout their lifetime. An exception to this are the high $\dot{m}>10^7$ systems where their short lifetimes (and therefore, a smaller acceleration region) restrict $E_{\rm max}<\varepsilon_{\rm min}$ and prevent any neutrino production.

The neutrino luminosities from the JTS-accelerated $p\gamma$ and $pp$ interactions are,
\begin{subequations} \label{eq:L_nu}
\begin{gather}
\begin{align}
    L_{\rm \nu}^{\rm p\gamma} & = & \frac{3}{8}f_{\rm p\gamma}L_{\rm p} & = \frac{27\pi}{64}f_{\rm p\gamma}f_{\rm E}R^2\rho_{\rm jts}(v_{\rm j}-v_{\rm fs})^3  \\
    L_{\rm \nu}^{\rm pp} & = & \frac{1}{2}f_{\rm pp}L_{\rm p} & = \frac{9\pi}{16}f_{\rm pp}f_{\rm E}R^2\rho_{\rm jts}(v_{\rm j}-v_{\rm fs})^3
\end{align}
,
\end{gather}
\end{subequations}
where the neutrino conversion efficiency factors are, 
\begin{subequations} \label{eq:f_p}
\begin{gather}
\begin{align}
    f_{\rm p\gamma}^{\rm (th,nth)} & = 1 - e^{-\tau_{\rm p\gamma}} = 1 - e^{-t_{\rm cross}^{\rm cone}/t_{\rm p\gamma}^{\rm (th,nth)}}  \\
    f_{\rm pp} & = 1 - e^{-\tau_{\rm pp}} = 1 - e^{-t_{\rm cross}^{\rm cone}/t_{\rm pp}} 
\end{align}
.
\end{gather}
\end{subequations}
The factor 3/8 in Eq.~\ref{eq:L_nu}(a) accounts for the fraction of the proton energy that goes into neutrino products \citep{Waxman&Bahcall97, Fang&Metzger17}, i.e., charged pions are produced in a $p\gamma$ interaction only half the time on average, and when a pion decays, $\sim$3/4 of its energy goes to neutrinos (see Eqs.~\ref{eq:pgamma} and \ref{eq:pion_decay}). The factor 1/2 in Eq.~\ref{eq:L_nu}(b) accounts for the fact that pions are produced with $\sim$2/3 probability in a $pp$ interaction (see Eq.~\ref{eq:pp}), and $\sim$3/4 of their energy is carried away by neutrinos. 
Finally, each hypernebula provides a total neutrino energy flux per flavor that follows the proton energy spectrum at a given time as,
\be \label{eq:neutrino_spectrum}
E^2_\nu \frac{dN_\nu}{dE_\nu} = \frac{E^2_{\rm p}}{2}\frac{dN_{\rm p}}{dE}\left[\frac{3}{4}f_{\rm p\gamma}^{\rm (th,nth)} + f_{\rm pp}\right].
\ee
Note that in the calculation above for the neutrino luminosity and its flux, we assume for simplicity that the opening angle of the JTS ($\theta_{\rm j}$) is comparable to that of the X-ray cone ($\theta_{\rm X}$; Eq.~\ref{eq:f_b}). On the other hand, one can expect a scenario where $\theta_{\rm X}>\theta_{\rm j}$, and that the precessing X-ray cone irradiates different patches of the JTS at different times. In such a case, the effective expansion time (E.g., see Eq.~\ref{eq:t_cross_cone}) could be lower by a factor $(\theta_{\rm X}/\theta_{\rm j})^2$. This ratio can be estimated through magnetohydrodynamic simulations of advection-dominated hyper-Eddington accretion disks, and is beyond the scope of this work.

\section{Observational Implications} \label{sec:observable_implications}
\subsection{Luminosity and volumetric rates} \label{subsec:luminosity_rates}

\begin{figure} 
\centering
\includegraphics[width=\linewidth]{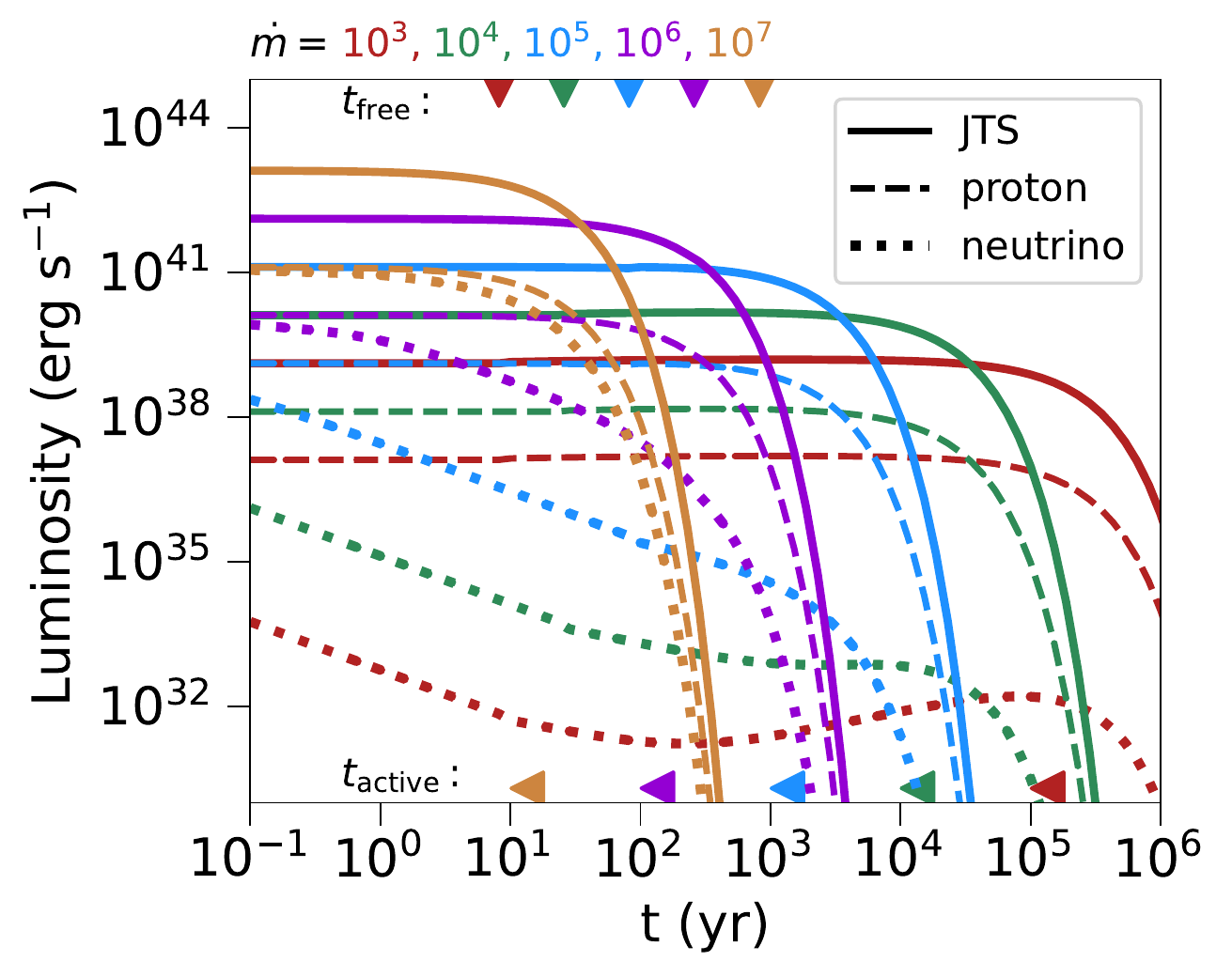}
\caption{Solid, dashed, and dotted curves show the temporal luminosity evolution of jet termination shock (Eq.~\ref{eq:L_s}), and the protons and neutrinos energized there (Eqs.~\ref{eq:f_E} and \ref{eq:neutrino_spectrum}), respectively. Different colors represent different accretion rates. The left-facing triangles in the lower frame represent the active duration of the engine ($t_{\rm active}$; Eq.~\ref{eq:t_active}) for different $\dot{m}$, and the bottom-facing triangles along the upper frame mark the end of the free expansion of the shell ($t_{\rm free}$; Eq.~\ref{eq:t_free})}
\label{fig:luminosities}
\end{figure}

Fig.~\ref{fig:luminosities} shows the combined $L_{\nu}^{\rm p\gamma}+L_{\nu}^{\rm pp}$ neutrino `light curves' for different $\dot{m}$. The neutrino luminosity closely follows the profile of the JTS and proton luminosity for $\dot{m}=10^7$ case, but it evolves as $\propto t^{-1}$ for $\dot{m}\lesssim10^6$ cases. For even lower $\dot{m}\lesssim10^4$ cases, the neutrino luminosity increases again before turning off: this is due to the onset of hadronic ($pp$) channels of neutrino production at later times. The peak neutrino luminosity is larger for higher $\dot{m}$: for the fiducial cases shown in Fig.~\ref{fig:luminosities}, the brightest emission ($L_{\nu}=10^{41}\,{\rm erg\,s^{-1}}$) is seen from $\dot{m}=10^7$ during the early moments of JTS onset. 

Whether neutrino emission can be observed from an individual hypernebulae, however, depends on the abundance of such sources in the universe, if the emission is nearly isotropic.  The volumetric rate of hypernebulae can be assumed to follow common envelope events involving a black hole/neutron star with donor stars more massive than $m_\star=10$. \cite{Schroder+20} find this frequency to be $\sim0.6\%$ core-collapse supernova rate, corresponding to a local-universe (redshift $z=0$) volumetric rate of ${\cal R}_0\sim10^3\,{\rm Gpc^{-3}\,yr^{-1}}$ \citep{Strolger+15,Vigna-Gomez+18}. This rate is also in agreement (within an order of magnitude) with that of engine-powered fast blue optical transients (FBOTs; ${\cal R}_0\sim 300\,{\rm Gpc^{-3}\,yr^{-1}}$) which could be failed common envelope events \citep{Coppejans+20, Ho+21b,Metzger22}, the occurrence rate of massive star binaries with mass-transfer rates $\dot{M}\gtrsim\dot{M}_{\rm Edd}$ (${\cal R}_0\sim 100\,{\rm Gpc^{-3}\,yr^{-1}}$; \citealt{Pavlovskii+17}), and $\sim1-10\%$ of all the luminous red novae merger events \citep{Karambelkar+22}. Assuming an active timescale of $\sim$10\,yr (for the neutrino-brightest $\dot{m}=10^7$ case), one active hypernebula can be expected within $50$\,Mpc. In the most optimistic scenario (considering $\eta\sim1$), the peak neutrino flux from a `local' ($\sim$50\,Mpc) source would be $10^{-10}\,{\rm GeV\,s^{-1}\,cm^{-2}\,sr^{-1}}$, suggesting that the prospects of detecting neutrinos from individual hypernebulae are not encouraging. 

However, if the emission is beamed, then the apparent neutrino luminosity of each of the individual sources would be up to $10^{48}\,\dot{m}_7\,{\rm erg\,s^{-1}}$ (also see Fig.~\ref{fig:luminosities}). Assuming a volumetric rate of ${\cal R}_0\sim 300\,{\rm Gpc^{-3}\,yr^{-1}}$ (corresponding to FBOTs), the apparent density of such sources would be $3\times10^{-5} \dot{m}_7^{-1}\,{\rm Gpc^{-3}\,yr^{-1}}$. Assuming a detection sensitivity of $\Phi_{\nu_\mu}\sim 3\times 10^{-12} \,{\rm TeV^{-1}\,cm^{-2}\,s^{-1}}$ at 1\,TeV \citep{Aartsen+20}, a source with an $E^{-2}$ spectrum between $10^3$ and $10^7$\,GeV (based on Fig.~\ref{fig:neutrino_flux}) can be observed up to a comoving distance of $\sim 8\, \dot{m}_7^{1/2}$\,Gpc, which corresponds to $\sim 0.015 \, \dot{m}_7^{1/2}$ source per year. Considering that the emission duration of a $\dot{m}=10^7$ source is only $\sim 10$\,yr, such a rate is consistent with the absence of a single hypernebula neutrino source detected by IceCube during its $\sim$10\,yr operation duration.

Recently, IceCube reported a 1\,TeV flux of $\simeq1.5\times10^{-7}\,{\rm GeV\,s^{-1}\,cm^{-2}}$ from NGC\,1068 \citep[at 14.4\,Mpc; ][]{IceCubeCollaboration_22}. The peak flux from a single hypernebula accreting at $\dot{m}=10^7$ situated at that distance ($\sim$10\% chance) would be $2.5\times10^{-9}\,{\rm GeV\,s^{-1}\,cm^{-2}}$ (see Fig.~\ref{fig:luminosities}). Therefore, the observed neutrino flux from NGC\,1068 could in principle, be explained by a single hypernebula if (1) the JTS is weakly magnetized $\sigma_{\rm j}\lesssim10^{-3}$, resulting in a larger $\epsilon_{\rm rel}\sim0.1$ \citep{Sironi&Spitkovsky11}, or (2) in a scenario where a cluster of hypernebulae migrate to the AGN core \citep{Secunda+20}, while the previous points (1) and (2) should not necessarily be satisfied in most of the hypernebulae lest they violate the constraints from the diffuse flux. Though, we note that other, more favorable, channels involving AGN coronae could well be operating \citep{Murase+20, Inoue+22, Eichmann+22}.

As far as the HE diffuse background neutrino flux is concerned, its overall normalization can be explained by non-relativistic shock-powered transients as long as they obey \citep{Fang+20},
\be \label{eq:transient_requirements}
\left(\frac{{\cal R}_0}{10^5\,{\rm Gpc^{-3}\,yr^{-1}}}\right)\left(\frac{E_{\rm tot}}{5\times10^{50}\,{\rm erg}}\right)\left(\frac{\epsilon_{\rm rel}}{1\%}\right)\sim 1,
\ee
where the total output energy is $E_{\rm tot}$, of which, a fraction $\epsilon_{\rm rel}$ is directed toward accelerating non-thermal ions (see Eq.~\ref{eq:f_E}).  While the volumetric rates of hypernebulae are smaller than the suggested value in Eq.~\ref{eq:transient_requirements} (which is normalized to that of core-collapse supernovae; \citealt{Briel+22}), their longer lifetimes, and steady injection of energy into their surroundings brings the total energy budget of any hypernebula (nearly independent of $\dot{m}$) to
\be
E_{\rm tot} \approx \int^{t_{\rm active}}L_{\rm w}dt \sim 5\times10^{52}\,{\rm erg}\left(\frac{M_\star}{30\,M_\odot}\right)v_{\rm w,9}^2.
\ee
This demonstrates that hypernebulae could in principle be a significant portion of the HE diffuse background neutrino flux. We note here that the total energy budget and the volumetric rates of hypernebulae are comparable to hypernovae, which have been considered as possible cosmic ray sources \citep{Chakraborti+11, Senno+15}. The decades- to centuries-long radio X-ray and other multi-wavelength counterparts of hypernebulae \citep{Sridhar&Metzger22} can be used to tell them apart from hypernovae through stacking analysis (akin to e.g., \citealt{Abbasi+23}).

Fig.~\ref{fig:luminosities} shows that although neutrinos are generated throughout the nebula's lifetime, the spectral properties constantly evolving as the hypernebula expands (see also Fig.~\ref{fig:protonspectra}). To calculate a representative neutrino flux for each hypernebula, we calculate the weighted-sum of the flux emitted by each hypernebula Eq.~\ref{eq:neutrino_spectrum} over its lifetime. The integrated observed neutrino flux contributed by all hypernebulae from different redshifts, taking into account the energy loss due to cosmological expansion can be written as,
\be \label{eq:neutrino_flux}
E_\nu^2\Phi_\nu(E_\nu) = \frac{{\cal R}_0}{4\pi}\int_0^6 dz \frac{c}{(1+z)^2H(z)}f(z)\left({E_\nu^\prime}^2\frac{dN_\nu}{dE_\nu^\prime}\right)(z), 
\ee
where $H(z)=H_0\left[\Omega_{\rm m}(1+z)^3+\Omega_\Lambda\right]^{1/2}$ is the Hubble constant at redshift $z$, where we take $H_0\simeq70\,{\rm km\,s^{-1}\,Mpc^{-1}}$, $\Omega_{\rm m}=0.315$, and $\Omega_\Lambda=0.7$ \citep[for a $\Lambda$CDM flat cosmology;][]{Planck_Collaboration20}. $E_\nu^\prime=E_\nu(1+z)$ is the redshifted neutrino energy, $f(z)={\cal R}(z)/{\cal R}_0$ captures the possibly different volumetric rates of the source at different redshifts. We assume ${\cal R}(z=0)\sim10^3\,{\rm Gpc^{-3}\,yr^{-1}}$, and that ${\cal R}$ follows star formation history. This lets us obtain $f(z)$ using the following empirical fit (reliable upto $z\sim6$) of \cite{Strolger+04},
\be\label{eq:SFH}
f(z) = \frac{a\left[t^be^{-t/c} + de^{d(t-t_0)/c}\right]}{de^{-dt_0/c}},
\ee
where $a=0.182$, $b=1.26$, $c=1.865$, $d=0.071$, $t=1/H_0(1+z)$ is the look-back time, and $t_0=1/H_0\sim14\,$Gyr is the age of the universe. Since the volumetric rate of hypernebulae scales as the star formation history ($f(z)$, which increases with redshift) the total number of hypernebulae contributing to the neutrino flux is $\sim10\times$ what one would get if a constant ${\cal R}={\cal R}_0$ is assumed.

\subsection{Diffuse Neutrino Flux} \label{subsec:neutrino_flux}

\begin{figure} 
\centering
\includegraphics[width=1\linewidth]{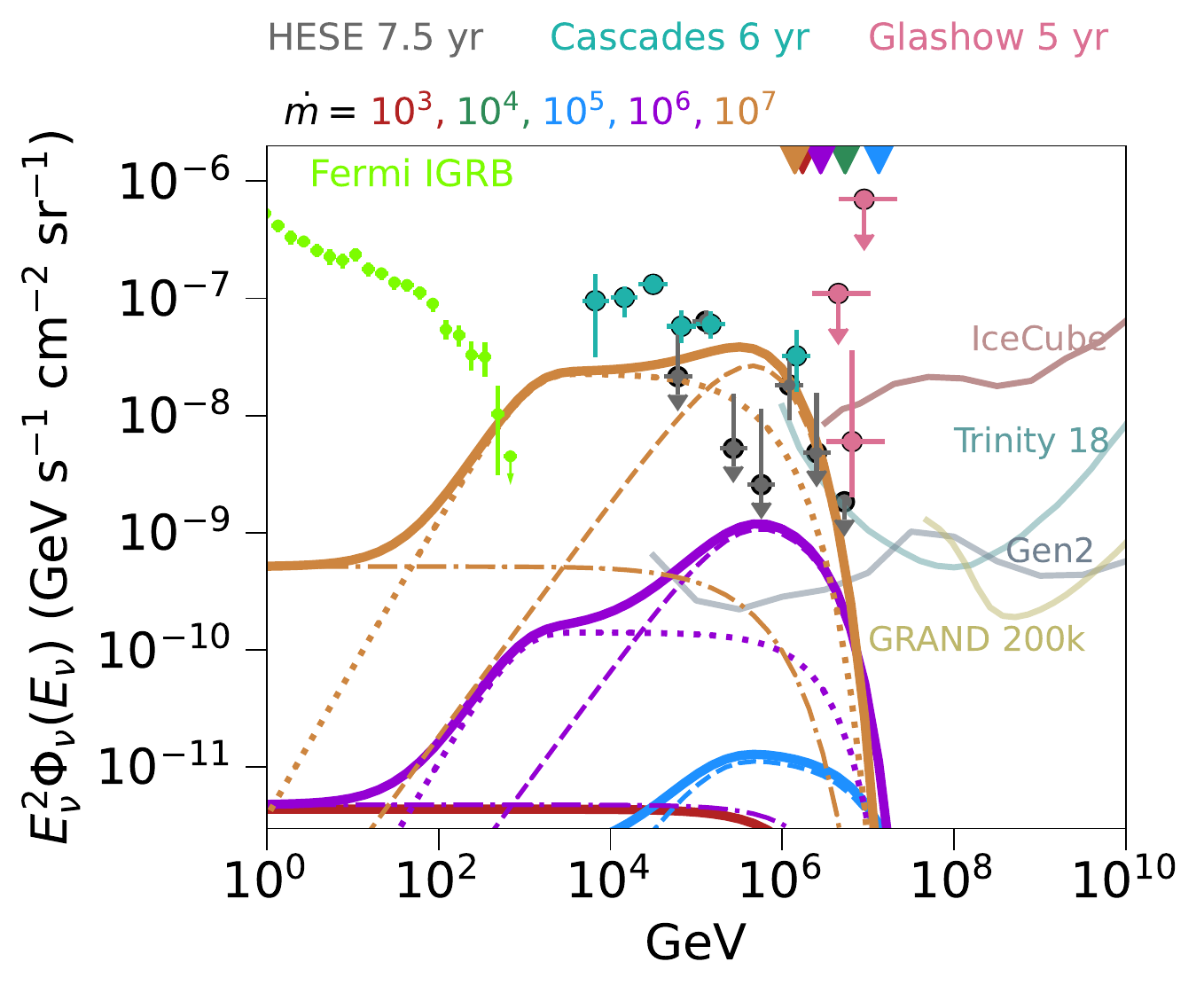}
\caption{All-flavor HE diffuse background neutrino spectra. The IceCube measurements shown are from the 7.5\,yr high-energy starting event sample \citep[grey markers;][]{IceCube_HESE+21}, 6\,yr high-energy cascades \citep[turquoise markers;][]{Aartsen+20}, and the 5\,yr Glashow resonance \citep[pink markers][]{IceCube_Collaboration+21}. For comparison, the \textit{Fermi} extra-galactic isotropic gamma-ray background is shown with green markers \citep{Ackermann+15}. Error bars and upper limits represent 68\% confidence intervals. The  10\,yr sensitivity limits of current (IceCube, in light maroon, is based on detected neutrino events; \citealt{Aartsen+13}) and future (Trinity with 18 telescopes: light green; \citealt{Nepomuk_Otte_19}, IceCube-Gen2 : grey; \citealt{IceCube-Gen2+14, Aartsen+21}, GRAND with 200,000 stations : light gold; \citealt{Alvarez-Muniz+20}) neutrino detectors are also provided for comparison. \textit{Left panel:} Solid curves are the total volume-integrated neutrino flux during the active lifetime of hypernebula obtained for models with different accretion rates (color-coded) but otherwise the same fiducial parameters ($v_{\rm w}=0.03c, v_{\rm j}=0.3c, n=10, \sigma_{\rm j}=0.1, \eta=0.5$; see \S\ref{sec:engine_properties} for more details). Dashed and dotted curves represent the contribution of photomesonic ($p\gamma$) interaction with thermal and non-thermal photons, respectively, and dash-dotted curves denote the contribution of hadronic interactions ($pp$). The maximum neutrino energies for different $\dot{m}$ are color-coded and marked as down-facing triangles along the top frame.
}
\label{fig:neutrino_flux}
\end{figure}

The left panel of Fig.~\ref{fig:neutrino_flux} shows the time-averaged, volume-integrated (Eq.~\ref{eq:neutrino_flux}) neutrino flux energy distribution for different mass-transfer rates $\dot{m}$ (denoted by different colors). The individual processes contributing to the total flux are shown with different linestyles i.e., the dominant $p\gamma$ photomesonic reactions with thermal and nonthermal background photons, and the highly sub-dominant hadronic $pp$ reactions, whose contribution to the neutrino flux is relevant only at lower-energies $<$100\,GeV. Also shown for comparison are the observed HE diffuse background neutrino flux \citep{Aartsen+20}, the \textit{Fermi}/LAT IGRB \citep{Ackermann+15} (circular markers with error bars), and the 10\,yr sensitivity limit of future high-energy and UHE neutrino observatories. 

The neutrino spectral peak is primarily determined by the $p\gamma$ interactions with the high-density thermal photons. The high-energy cut-off is set by the system parameters, independent of the neutrino production channel (Eq.~\ref{eq:E_max} and Fig.~\ref{fig:energies}). On the other hand, the low-energy turnovers are set by the three different neutrino production channels: the turnovers at $\sim$100\,TeV and $\sim$10\,TeV are due to the $p\gamma$ interactions with the thermal and nonthermal photons, respectively (see discussion following Eq.~\ref{eq:varepsilon_min}). The lowest-energy cut-off at $\sim$135\,MeV (below the shown energy range of Fig.~\ref{fig:neutrino_flux}) is set by the photohadronic component of the spectrum, at the pion creation threshold energy. The locations of these low-energy turnovers notably depend only weakly on the hypernebula parameters, making them predictive of their neutrino emission (as we demonstrate in Appendix~\ref{appendix:parameter_exploration}).

The left panel of Fig.~\ref{fig:neutrino_flux} shows that if all binaries generate conditions following our high mass-transfer rate model $\dot{m}\sim10^7$ (brown curve; with otherwise the same fiducial parameters mentioned in \S\ref{sec:engine_properties}) such systems can in principle explain the entirety of the diffuse 10\,TeV to PeV neutrino flux (the chosen parameter combination underestimates the 10\,TeV neutrinos, which shall be addressed below with different model parameters).  By contrast, the $\dot{m}\sim10^6$ and $\dot{m}\sim10^5$ models, acting alone, can supply at most $\sim1\%$ and $\sim0.01\%$ of the observed flux, thus ruling out $\dot{m}\ll10^6$ binaries as significant contributors. Although not visibly shown, the flux in the $\dot{m}\sim10^3$ case reaches values as high as $\gtrsim10^{-12}\,{\rm GeV\,s^{-1}\,cm^{-2}\,sr^{-1}}$---comparable to $\dot{m}\sim10^4$ case---because the long active time of the jet allows $pp$ channel to dominate the waning $p\gamma$ channel for neutrino production. 

Appendix~\ref{appendix:parameter_exploration} explores a range of models in which the parameters are varied about the fiducial model choices.  This exploration yields a `best-fit' model to the diffuse neutrino flux, for the hypernebula parameters $\{\dot{m}=10^7, v_{\rm w}/c=0.07, v_{\rm j}/c=0.17, \eta=1, n=1, \sigma_{\rm j}=1$, $q=2\}$. The robust low-energy turnover feature of the model is consistent with the flat/mildly rising nature of the observed neutrino spectrum from 10--50\,TeV, followed by a broken power-law at $\gtrsim$\,PeV.  Though we present this fit as a proof of principle, we do not intend to suggest that the entire observed diffuse flux is produced by a special population hypernebula; instead, we offer it as a representative of the parameter space around which hypernebulae should occupy, to contribute appreciably to the diffuse flux. Indeed, certain portions of hypernebula parameter space are ruled out because they would overproduce the number of synchrotron-emitting nebula observed by VLASS (e.g., for $v_{\rm j}/c \ngtr 0.5$; \citealt{Sridhar&Metzger22}); our model to the neutrino diffuse background is consistent with this constraint. Our parameter exploration in Appendix~\ref{appendix:parameter_exploration} reveals that different parameter combinations can extend the flux cutoff to higher energies, making such systems candidate UHE neutrino sources for upcoming neutrino observatories such as IceCube-Gen2 \citep{IceCube-Gen2+14, Aartsen+21}, Trinity \citep{Nepomuk_Otte_19}, and GRAND \citep{Alvarez-Muniz+20}. UHE experiments with a sensitivity window well above $\sim$~EeV are unlikely to detect these sources. 

\begin{figure} 
\centering
\includegraphics[width=1.0\linewidth]{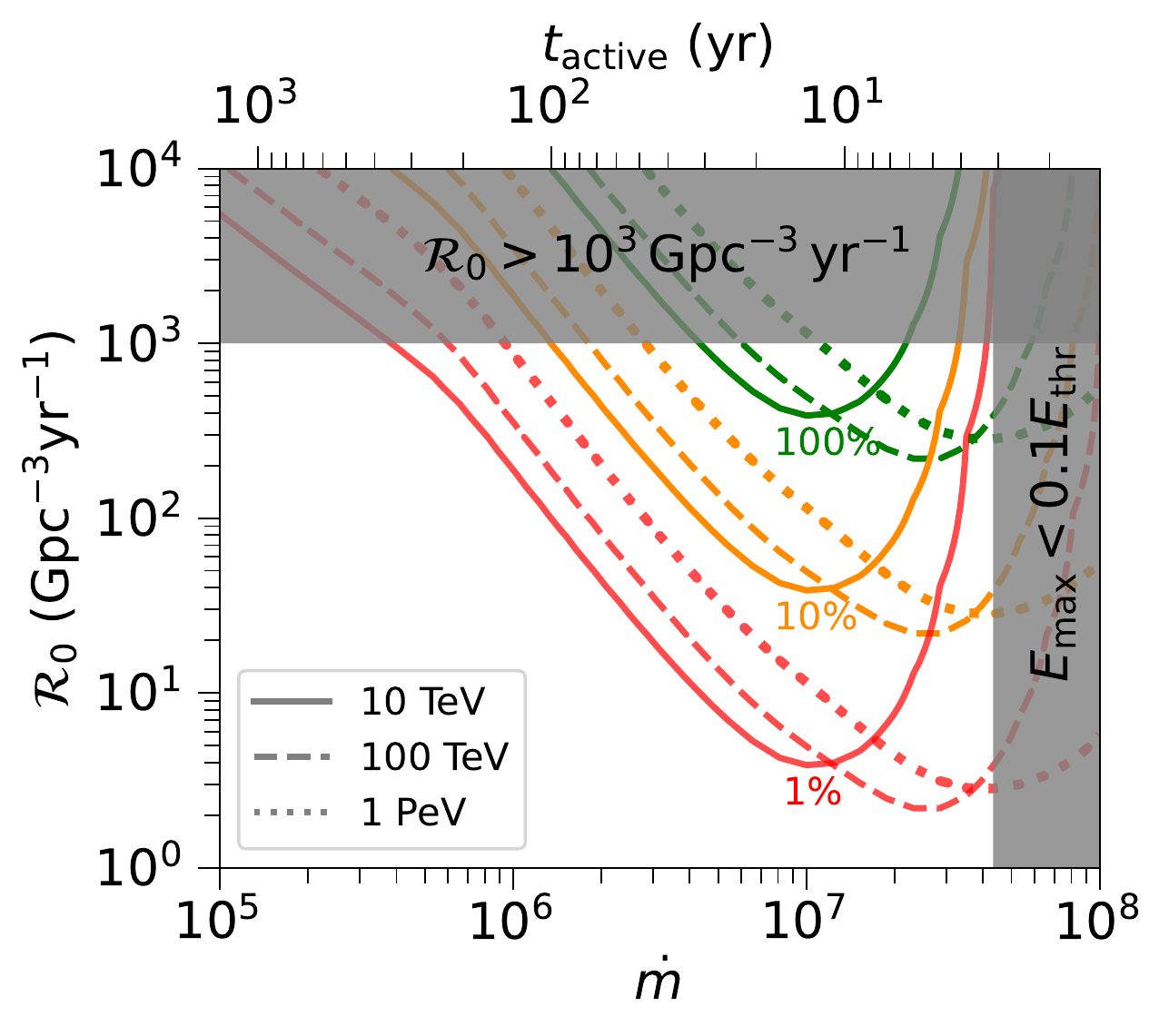}
\caption{Neutrino flux for a range of volumetric rates ${\cal R}_0$ (y-axis) from accretion-powered-hypernebulae with different $\dot{m}$ (bottom x-axis), with other parameters following our `best fit' (see \S\ref{subsec:neutrino_flux}, and Appendix~\ref{appendix:parameter_exploration}); the top x-axis denotes the active timescale (Eq.~\ref{eq:t_active}) for a given $\dot{m}$. The solid, dashed and dotted curves represent the model fluxes in 10\,TeV, 100\,TeV, and 1\,PeV, normalized to the observed fluxes of $10^{-7},~4\times10^{-8},~{\rm and}~2\times10^{-8}\,{\rm GeV\,s^{-1}\,cm^{-2}\,sr^{-1}}$, respectively. Red, yellow, and green contours denote the parameter space that can supply 1\%, 10\%, and 100\% of the observed flux. The top grey region denotes the parameter space excluded based on the volumetric rate constraint of ${\cal R}_0\lesssim10^3\,{\rm Gpc^{-3}\,yr^{-3}}$ set by the rate of common envelope events involving compact objects \citep{Schroder+20}. The right grey region is excluded because extreme $\dot{m}$ events fail to produce neutrinos as their maximum proton energies ($E_{\rm max}$; Eq.~\ref{eq:E_max}) is less than the minimum required proton energy to produce neutrinos upon interaction with even the higher-energy nonthermal photons (Eq.~\ref{eq:E_thr}).
}
\label{fig:parameter_space}
\end{figure}

Above we have made the assumption that the entire population of hypernebulae is characterized by a single $\dot{m}$, which is an obvious oversimplification.  In reality, systems with different $\dot{m}$ will be sampled by nature at different volumetric rates  ${\cal R}_0$, and even a single binary is likely to span different values of $\dot{m}$ throughout its mass-transfer evolution.  We explore this degeneracy further in Fig.~\ref{fig:parameter_space}, where the different contours in the ${\cal R}_0-\dot{m}$ space indicate the model neutrino flux at different energies (as a fraction of the observed diffuse flux). We see that sources with $\dot{m}<4\times10^5$ cannot explain even 1\% of the diffuse flux for any realistic occurrence rate ${\cal R}_0\lesssim10^3\,{\rm Gpc^{-3}\,yr^{-1}}$ (red contours). By contrast, hypernebula models with $\dot{m}\gtrsim10^6$ can explain a significant fraction ($\gtrsim10\%$; yellow contours) of the observed 10\,TeV--1\,PeV flux even with a lower local volumetric rate of ${\cal R}_0={\rm few}\times100\,{\rm Gpc^{-3}\,yr^{-1}}$. Only $\dot{m}\sim10^7$ models are able to explain all of the observed flux (green contours). This is relevant because the lives of many binaries may culminate in a highly unstable mass transfer stage (with $\dot{m}\gtrsim10^7$), before the compact object plunges into the envelope of the donor star.  On the other hand, in a conservative scenario where hypernebulae accrete at a constant rate throughout their lives, with volumetric rates uniformly distributed across a range of mass-transfer rates $10^3\lesssim\dot{m}\lesssim10^7$, then their total contribution hypernebulae to the  10\,TeV--1\,PeV diffuse flux would be $\sim25\%\,({\cal R}_{0}/10^3\,{\rm Gpc^{-3}\,yr^{-1}})$, and will be dominated by the $\dot{m}\sim10^7$ cases. 

Throughout our analysis, we do not consider sources accreting at rates $\dot{m}\gg10^7$ because, their short lifetimes would preclude any proton from being accelerated to energies $\gtrsim0.1E_{\rm thr}$ (Eq.~\ref{eq:E_thr}) to engage in neutrino production through photomesonic channels (the factor of 0.1 accounts for $p\gamma$ interaction with the higher energy nonthermal photons, which requires a smaller $E_{\rm thr}$; vertical grey region in Fig.~\ref{fig:parameter_space}).  In the following section, we consider the gamma-ray properties of the more relevant $\dot{m}=10^7$ scenario.

\subsection{$\gamma$-ray attenuation} \label{subsec:gamma-ray_attenuation}

\begin{figure} 
\centering
\includegraphics[width=1.0\linewidth]{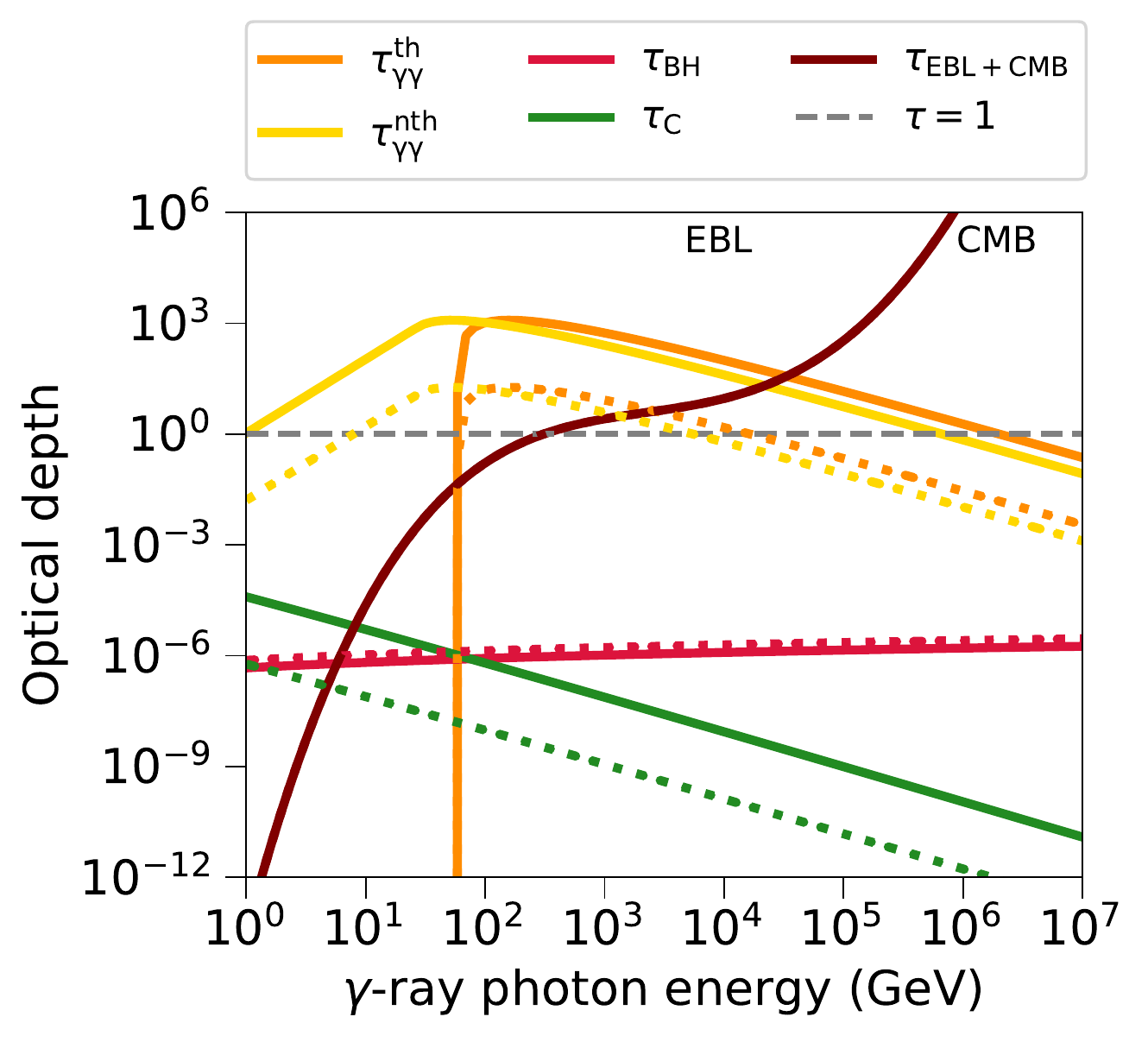}
\caption{Optical depths of various interactions to gamma-ray photons with different energies. Solid and dotted curves are at times $t=t_{\rm active}$ (Eq.~\ref{eq:t_active}) and $t=t_{\rm free}$ (Eq.~\ref{eq:t_free}), respectively, for our fiducial model, but with the more relevant $\dot{m}=10^7$ case. Orange and yellow curves represent $\gamma\gamma\rightarrow e^-e^+$ pair production with the accretion disk thermal (Eq.~\ref{eq:tau_gg_th}) and nonthermal Comptonized photons (Eq.~\ref{eq:tau_gg_nonth}), respectively; Bethe-Heitler interaction with the ions in the ejecta is represented by red curves (Eq.~\ref{eq:tau_BH}); green curves represent the Compton down-scattering in the Klein-Nishina regime with electrons in the ejecta (Eq.~\ref{eq:tau_C}), and the interaction with extragalactic background light (EBL) and cosmic microwave background (CMB) radiation, for source at a redshift $z=0.2$, is represented by maroon curves: the energies where EBL and CMB interactions dominate are noted near the top frame of the figure.}
\label{fig:gammaray_escape}
\end{figure}

An inevitable byproduct of neutrino production are gamma-rays (from neutral pion decay; see Eq.~\ref{eq:pion_decay}a), with flux and spectral properties similar to those of neutrinos. 
The Fermi IGRB observations indicate a lower gamma-ray flux than neutrinos \citep{Ackermann+15} suggesting that the primary sources of HE diffuse background neutrino should be faint gamma-ray emitters. In this section, we discuss the opacity of hypernebula to the gamma-rays generated along with the neutrinos, and the impact of the intervening medium as the gamma-rays propagate to Earth. The emitted gamma-rays can be attenuated and reprocessed at least by the following three intrinsic processes, and Fig.~\ref{fig:gammaray_escape} shows some of the relevant optical depths at times $t=t_{\rm free}$ and $t=t_{\rm active}$.

(1) \textit{Compton down-scattering:} 
The gamma-ray photons can lose energy and be attenuated by scattering off the electrons in the ejecta. The effective optical depth for the photons to lose most of their initial energy is \citep{Vurm&Metzger21},
\begin{align} \label{eq:tau_C}
\tau_{\rm C} & \simeq \frac{\dot{x}_{\rm C}}{c}\frac{R}{c}(1 + \tau_{\rm KN}) \approx \tau_{\rm T}\frac{\ln(1+x)}{(1+3x)}\left[1+\frac{3\tau_{\rm T}}{8x}\ln\left(1+\frac{8x}{3}\right)\right] \nonumber \\
 & \approx 10^{-9}\,\ln\left(\frac{x}{10^6}\right)\left(\frac{x}{10^6}\right)^{-1} \nonumber \\
 &\times
    \begin{cases}
       \dot{m}_{5} v_{\rm w,9}^{-2}\left(\frac{t}{80\,{\rm yr}}\right)^{-2} & (t<t_{\rm free}) \\
       \dot{m}_{5}^{1/5} v_{\rm w,9}^{2/5}n_{\rm 1}^{4/5}\left(\frac{t}{80\,{\rm yr}}\right)^{3/5} & (t>t_{\rm free})
    \end{cases}
,
\end{align}
where $\tau_{\rm KN}$ is the Klein-Nishina optical depth for gamma-ray photons with dimensionless energies $x\equiv E_{\gamma}/m_{\rm e}c^2\gg1$ \citep{Klein&Nishina29}; the factor $1+\tau_{\rm KN}$ accounts for the diffusion time of the photons through the ejecta. $\tau_{\rm T}\sim n_{\rm e}\sigma_{\rm T}R$ is the Thomson optical depth with a scattering cross section $\sigma_{\rm T}=6.6\times10^{-25}\,{\rm cm}^{-2}$ for an electron number density $n_{\rm e}\sim n_{\rm p}$, and the particle energy loss rate is $\dot{x}_{\rm C}/x\approx c\sigma_{\rm T}n_{\rm e}\ln(1+x)/(1+3x)$. Clearly, even the low-energy ($E_\gamma\sim1$\,GeV) gamma-ray photons do not suffer any losses due to Compton scattering, and the higher energy photons only experience even lesser attenuation due to this process.

(2) \textit{Bethe-Heitler process:}  gamma-rays can be attenuated due to pair production upon interaction with relativistic ions in the ejecta \citep{Bethe&Heitler34}. The corresponding optical depth is given by \citep{Chodorowski+92},
\begin{align} \label{eq:tau_BH}
\tau_{\rm BH} = n_{\rm p}\sigma_{\rm BH}R & \approx 10^{-5}\ln\left(\frac{x}{10^6}\right) \nonumber \\
& \times
    \begin{cases}
       \dot{m}_{5} v_{\rm w,9}^{-2}\left(\frac{t}{80\,{\rm yr}}\right)^{-2} & (t<t_{\rm free}) \\
       \dot{m}_{5}^{1/5} v_{\rm w,9}^{2/5}n_{\rm 1}^{4/5}\left(\frac{t}{80\,{\rm yr}}\right)^{3/5} & (t>t_{\rm free})
    \end{cases}
.
\end{align}
The scattering cross-section can be approximated as \cite{Zdziarski&Svensson89}, 
\be \label{eq:sigma_BH}
\sigma_{\rm BH} \simeq \frac{3}{8\pi}\alpha\sigma_{\rm T}Z^2\left[\frac{28}{9}\ln(2x)-\frac{218}{27}\right] 
\ee
where $\alpha=1/137$ is the fine structure constant, $Z$ is the atomic charge of the nuclei (we assume $Z \simeq 1$ for hydrogen-dominated composition)\footnote{However, note that if the donor star has lost a significant fraction of its envelope, then the nebula material might possess a heavier-than-solar composition. Nonetheless, this will not qualitatively change our conclusion that Bethe-Heitler absorption is negligible.}. We can see from Fig.~\ref{fig:gammaray_escape} that $\tau_{\rm BH}$ barely depends upon $E_{\gamma}$, and the effect of Bethe-Heitler process is negligible in attenuating the emitted gamma-rays. 

(3) \textit{$\gamma\gamma\rightarrow e^+e^-$ pair production:} Gamma-rays pair produce upon interacting with ambient low-energy photons \citep{Breit&Wheeler34}. Assuming for simplicity, a monochromatic isotropic thermal ambient UV/X-ray photon field (with an energy given by Eq.~\ref{eq:kBTeff}), the energy of the gamma-ray photon $E_{\gamma}$ at which the pair production cross-section $\sigma_{\gamma\gamma}$ peaks is, 
\be \label{eq:E_gamma_pk}
E_{\gamma, \rm pk} = \frac{4 m_{\rm e}^2c^4}{k_{\rm B}T_{\rm eff}} \approx60\,{\rm GeV}(1 + \ln \dot{m}_{7})^{1/4}\dot{m}_{7}^{1/4} \left(\frac{v_{\rm j}}{0.3\,c}\right)^{-1/2}.
\ee
The scattering cross-section can be taken as a smoothly connected piece-wise function such that \citep{Svensson87}, 
\be \label{eq:sigma_gg}
\sigma_{\gamma\gamma}\sim \frac{3\sigma_{\rm T}}{16} \times
    \begin{cases}
    0 & (E_{\gamma}<E_{\gamma, \rm pk}/4)  \\
    1 & (E_{\gamma}=E_{\gamma, \rm pk})  \\
    \ln(x)/x & (E_{\gamma}>E_{\gamma,\rm pk})
    \end{cases}
    .
\ee
With this, the optical depth to $\gamma\gamma$ pair production with thermal photons can be estimated as,
\begin{align} \label{eq:tau_gg_th}
\tau_{\rm \gamma\gamma}^{\rm th} &= n_{\rm X}^{\rm th}\sigma_{\rm \gamma\gamma}R \approx 13\ln\left(\frac{x}{10^6}\right)\left(\frac{x}{10^6}\right)^{-1}(1 + \ln \dot{m}_{7})^{5/4}\dot{m}_{7}^{5/4} \nonumber \\
& \times\left(\frac{v_{\rm j}}{0.3\,c}\right)^{-1/2} \times
    \begin{cases}
    v_{\rm w,9}^{-1}\left(\frac{t}{80\,{\rm yr}}\right)^{-1} & (t<t_{\rm free})\\
    \dot{m}_{7}^{-1/5}v_{\rm w,9}^{-2/5}n_{\rm 1}^{1/5}\left(\frac{t}{80\,{\rm yr}}\right)^{3/5} & (t>t_{\rm free})
    \end{cases}
    .
\end{align}
Likewise, one can also define an optical depth for $\gamma\gamma$ pair production with nonthermal photons (see \S\ref{subsec:background_radiation}(2)), 
\be \label{eq:tau_gg_nonth}
\tau_{\gamma\gamma}^{\rm nth}=n_{\rm X}^{\rm th}\left(\frac{\epsilon}{\epsilon_{\rm pk,th}^{\rm eff}}\right)^{-2}\sigma_{\gamma\gamma}[\epsilon]R,
\ee
where $\sigma_{\gamma\gamma}[\epsilon]$ is constructed by taking $E_{\rm \gamma,pk}=4m_{\rm e}^2c^4/\epsilon$ for individual photons with energy $\epsilon$ in the nonthermal powerlaw distribution. Fig.~\ref{fig:gammaray_escape} shows that gamma-ray photons with energies in the range 1\,GeV $\sim$1\,PeV will be strongly attenuated within the hypernebula by $\gamma\gamma$ pair production: specifically, with interactions dominated by nonthermal photons in the energy range 1\,GeV-60\,GeV, and by thermal photons in the energy range 60\,GeV-1\,PeV. Intrinsic attenuation of $>$\,PeV gamma-rays is in principle possible by their interaction with the radio synchrotron photons. However, their small number density $n_{\rm GHz}\lesssim10^{-4}\,{\rm cm}^{-3}$ \citep{Sridhar&Metzger22} would imply a $\gamma\gamma$ pair production optical depth of $\tau_{\gamma\gamma}^{\rm GHz}\ll1$, and the higher energy gamma-rays ($E_\gamma\gtrsim$1\,PeV) would still escape freely. For our fiducial choice of parameters, as $\tau_{\rm C}\ll1$ (see Fig.~\ref{fig:gammaray_escape}),  no gamma-ray cascades would be generated in hypernebulae.

The escaping $>$PeV gamma-ray photons will interact with the Extragalactic Background Light (EBL) and the Cosmic Microwave Background (CMB) radiation on their way to Earth. The EBL is composed of the UV-Optical-IR photons ($0.1 - 1000\,\mu{\rm m}$) emitted by stars and galaxies since the epoch of reionization, which can pair produce upon interaction with gamma-rays with $E_\gamma\gtrsim$10\,TeV \citep[][see also Eq.~\ref{eq:E_gamma_pk}]{Nikishov1961, Gould&Schreder67a, Gould&Schreder67b}. 
In Fig.~\ref{fig:gammaray_escape}, we show the optical depth of EBL and CMB photons \citep{Stecker+06} to gamma-rays, and show that all the escaping gamma-rays from a hypernebula---which happen to have energies $E_\gamma>$PeV---will be attenuated. At different locations, the non-linear dynamics of the pair production involving processes such as inverse-Compton scattering ($e\gamma\rightarrow e\gamma$) and triple pair production process ($e\gamma\rightarrow ee^+e^-$) can initiate electromagnetic cascades until the energy of the later generation of particles falls below the threshold to sustain these reactions. As a result, the injected PeV gamma-rays could eventually appear at 10-100~GeV, with a spectral shape independent of the injected gamma-ray injection spectrum and the target soft photon spectrum \citep{Berezinskii&Smirnov75} and a flux below the {\it Fermi} IGRB \citep{Fang+22}.

\section{Conclusions} \label{sec:conclusion}

We have developed a model for high-energy neutrino emission from `hypernebulae'. A hypernebula is a new class of spatially compact ($\lesssim$pc), decades-millenia long transient `bubble' inflated by hyper-Eddington accredion disk winds that are more powerful than the typical Ultra Luminous X-ray sources \citep{Sridhar&Metzger22}. The rapid runaway mass transfer ($\dot{m}\sim10^3-10^7$) onto a stellar mass compact object (black hole or a neutron star) required to inflate hypernebulae is naturally expected from an evolved companion star (e.g., imminent giant) leading up to a common-envelope event.  For the purpose of our `proof-of-concept' calculations, we consider a fiducial binary containing a massive star ($M_\star=30\,M_\odot$) accreting onto a black hole of fixed mass ($M_\bullet=10\,M_\odot$), whose active duration is varied by varying the accretion rate ($10^3<\dot{m}<10^7$). The intrinsic properties of the hypernebula, and the observables of the neutrino emission (light curves, energy spectra, volume-integrated observed flux) are continuously evaluated across the free-expansion, deceleration, and post-active phases, as a function of the slow, wide-angled disk wind ($\dot{M}\sim\dot{M}_{\rm w}, v_{\rm w}$), and the fast, collimated disk wind/jet ($\eta=\dot{M}_{\rm j}/\dot{M}, v_{\rm j}, \sigma_{\rm j}$) properties. The take-away results of this paper are as follows:

    (1) The primary source of neutrino production in this model is the adiabatic fast wind/jet termination shock, where the shock-accelerated protons produce pions (that eventually decay into neutrinos) upon their photohadronic ($p\gamma$) interactions with the thermal and nonthermal X-ray photons from the accretion disk. The hadronic ($pp$) interactions play a sub-dominant role except during the late stages of a hypernebula's lifetime, particularly for low $\dot{m}$ systems.
    
    (2) The highest energy neutrinos are emitted during the end of a hypernebula's lifetime (when they are the largest in size). The maximum energy is found to be $\sim$300\,PeV for $\dot{m}=10^5$ case. UHE neutrinos are highly sought-after over the past half-a-century, and are likely to be discovered within the next couple of decades by upcoming neutrino observatories \citep{BrancoValera+22} such as CHANT \citep{Neronov+17}, Trinity \citep{Nepomuk_Otte_19}, GRAND \citep{Alvarez-Muniz+20}, IceCube-Gen2 \citep{IceCube-Gen2+14, Aartsen+21}, Askaryan Radio Array \citep{Ara_Collaboration+11}, and the ARIANNA detectors \citep{Barwick+15}. Hypernebulae are promising candidates for diffuse extragalactic ($\ll$EeV) UHE neutrinos.

    (3) A distinct feature of hypernebulae as neutrino sources is their property of robust minimum neutrino energy, which is set solely by the steady background photon energies: the thermal photons set the prominent $\sim$100\,TeV low-energy spectral turnover, the nonthermal photons introduce a turnover at $\sim$10\,TeV, and the $pp$ interactions set the lowest-energy turnover at 135\,MeV. IceCube-DeepCore \citep{Abbasi+12}, KM3NeT \citep[being built in the Mediterranean sea;]{Adrian-Martinez+16, Fermani20} and the Gigaton Volume Detector \citep[in lake Baikal;][]{Allakhverdyan+21} may extend and improve the current neutrino sensitivities to sub-TeV regime. These facilities will be able to test the hypernebula model by better characterizing the TeV and sub-TeV spectrum of the HE diffuse background neutrinos.

    (4) Assuming that the volumetric rates of hypernebulae follow common envelope events involving compact objects, an engine accreting at $\dot{m}=10^7$ could comfortably explain $\sim$100\% of the HE diffuse background neutrino flux. In a more realistic scenario, if the volumetric rate of hypernebulae were to be uniformly distributed among different accretion rate engines $10^3<\dot{m}<10^7$, hypernebulae could explain (depending on the mass-transfer rate) $\sim25\%\,({\cal R}_{0}/10^3\,{\rm Gpc^{-3}\,yr^{-1}})$ of the observed neutrino flux.

    (5) The gamma-rays that are inevitably produced along with neutrinos are attenuated within the hypernebulae primarily by the Breit-Wheeler $\gamma\gamma\rightarrow e^+e^-$ annihilation process. The gamma-ray photons in the energy range $1<E_\gamma/{\rm GeV}<60$ are attenuated by nonthermal background photons and the higher energy gamma-rays $60<E_\gamma/{\rm GeV}<10^6$ are attenuated by thermal background photons. The escaping $>$PeV gamma-rays will be attenuated by their interaction with the extragalactic background light. Therefore, hypernebulae are gamma-ray faint sources of neutrino emission, thus satisfying the isotropic gamma-ray background constraints of \textit{Fermi}/LAT.

Finally, we would like to emphasize that hypernebulae are radio-synchrotron-bright objects that can be detected as point sources with all-sky radio surveys such as the FIRST survey \citep[Faint Images of the Radio Sky at Twenty cm;][]{Becker+95} or VLASS \citep{Lacy+20}. A detailed study of the electromagnetic counterparts to hypernebulae---including light curves, spectra---can be found in \cite{Sridhar&Metzger22}. A population of FRBs could be powered by flares along hyper-Eddington accretion disk jets \citep{Sridhar+21b, Bhandari+22}. Recently, \cite{Abbasi+22b} presented the search results for spatially- and temporally-coincident neutrino emission from FRBs with IceCube Cascade data, and found no significant clustering around the short-lived and beamed FRB event. This result is in agreement with theoretical predictions \citep{Metzger+20}. However, in a scenario of accretion-powered FRBs, the hypernebula surrounding the central FRB engine could serve as a persistent radio, and neutrino source, thus also contributing to the HE diffuse background neutrino flux.

\section{Acknowledgement} \label{sec:acknowledgement}

N.S. acknowledges the hospitality of the Department of Astronomy at Cornell University, and the Center for Cosmology and AstroParticle Physics at The Ohio State University for their generous hospitality, where parts of this work were conceived. NS would like to thank John Beacom, Damiano Fiorillo, and Todd Thompson for fruitful discussions, and thank Stephanie Wissel for sharing the details on various Neutrino experiments' sensitivities. N.S. acknowledges the support from NASA (grant number 80NSSC22K0332), NASA FINESST (grant number 80NSSC22K1597), and Columbia University Dean's fellowship.  B.D.M. acknowledges support from the National Science Foundation (AST-2009255).  K.F. is supported by the Office of the Vice Chancellor for Research and Graduate Education at the University of Wisconsin-Madison with funding from the Wisconsin Alumni Research Foundation. K.F. acknowledges support from NASA (80NSSC22K1584) and National Science Foundation (PHY-2110821).

\appendix

\section{Timescales}
A list of all the relevant timescales of the model introduced in this paper, and their definitions, is provided in Table~\ref{tab:timescales}.

\begin{deluxetable}{ccl} \label{tab:timescales}
\tablenum{1}
\tablecaption{Model timescales.}
\tablewidth{0pt}
\tablehead{
\colhead{Variable} & \colhead{C.f.} & \colhead{Definition}
}
\startdata
$t_{\rm cross}$ & Fig.~\ref{fig:timescales} & Light-crossing timescale of the hypernebula.\\
$t_{\rm active}$ & Eq.~\ref{eq:t_active} & Active duration of the dynamically unstable accretion phase at a mass transfer rate of $\dot{M}$.\\
$t_{\rm free}$ & Eq.~\ref{eq:t_free} & Free-expansion timescale beyond which the wind ejecta shell starts decelerating.\\
$t_{\rm exp}$ & Eq.~\ref{eq:t_exp} & Expansion timescale of the shocked gas.\\
$t_{\rm cool}$ & Eq.~\ref{eq:t_cool} & Cooling timescale of the shock-heated gas.\\
$t_{\rm rad}^{\rm fs}$ & Eq.~\ref{eq:f_rad} & The time after which the forward shock becomes radiative.\\
$t_{\rm cross}^{\rm cone}$ & Eq.~\ref{eq:t_cross_cone} & The time particles spend within the solid angle of background X-ray photon field.\\
$t_{\rm p\gamma}^{\rm (th,nth)}$ & Eq.~\ref{eq:t_pgamma} & Cooling time due to photomeson interaction between protons and background thermal and nonthermal photon fields. \\
$t_{\rm pp}$ & Eq.~\ref{eq:t_pp} & Hadronuclear interaction timescale. \\
$t_{\pi,\rm cre}$ & Eq.~\ref{eq:t_pion} & Pion creation timescale. \\
$t_{\rm p,syn}$ & Eq.~\ref{eq:t_psyn} & Proton synchrotron cooling timescale. \\
$\tau_{\pi,\mu}$ & Eq.~\ref{eq:Esyn_pimu} & Decay times of pions and muons. \\
$t_{\rm acc}$    & Eq.~\ref{eq:t_acc} & Diffusive shock acceleration timescale.\\
\enddata
\end{deluxetable}

\section{Proton energy spectra}

\begin{figure} 
\centering
\includegraphics[width=0.45\linewidth]{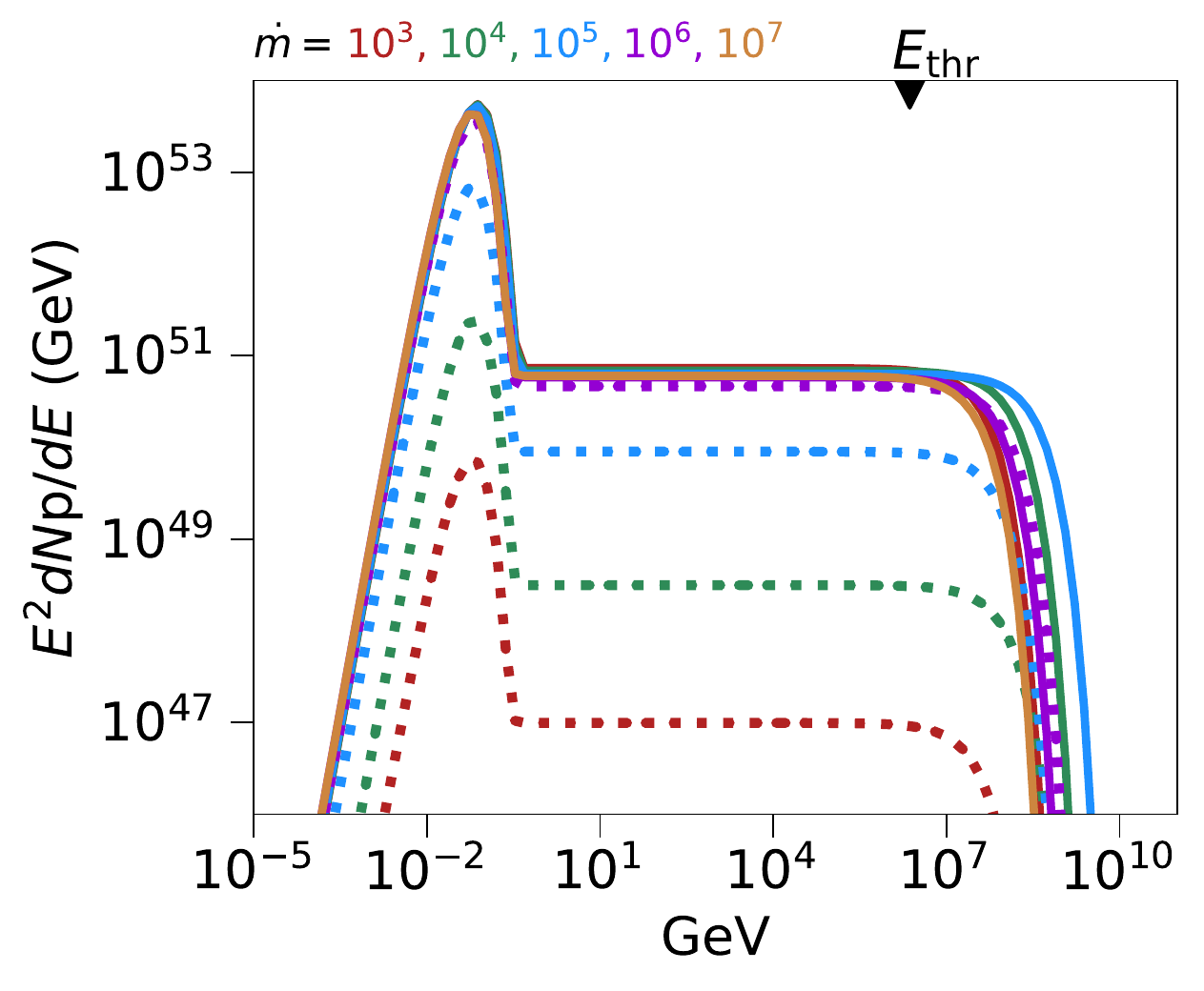}
\caption{Energy spectra of protons energized at the JTS. Solid curves are at $t=t_{\rm active}$ (Eq.~\ref{eq:t_active}), and the dotted curves are at an earlier time, $t=t_{\rm free}$ (Eq.~\ref{eq:t_free}). Different colors indicate different accretion rates. The threshold energy to produce neutrinos via photohadronic interaction with thermal background photons ($E_{\rm thr}$; Eq.~\ref{eq:E_thr}) is indicated along the top frame of the figure.}
\label{fig:protonspectra}
\end{figure}

The temporal evolution of the JTS-energized proton energy spectra is shown in Fig.~\ref{fig:protonspectra}, for different $\dot{m}$. The thermal population of protons occupies a distinct hump in the energy spectrum at $E\sim3\bar{E}\sim10^{-2}$\,GeV (see Eq.~\ref{eq:mean_temp}). At $t \sim t_{\rm free}$ (dotted curves), both the spectral high-energy cutoff, $E_{\rm max}(t)$, and the spectral normalization increase with $\dot{m}$. However, at later times $t \sim t_{\rm active}$ (solid curves), the protons accelerated by the higher-$\dot{m}$ engines are less energetic than the lower-$\dot{m}$ cases given the shorter lifetimes of the former, and the resulting size-limited energization constraints \citep{Hillas_84}. This reduces the high-energy spectral cut-off for high-$\dot{m}$ cases while retaining a constant $E_{\rm thr}$. This limits neutrino production only to cases with $\dot{m}\lesssim{\rm few}\times10^7$ (see also Fig.~\ref{fig:parameter_space}).

\section{Hypernebula parameter exploration} \label{appendix:parameter_exploration}

\begin{figure} 
\centering
\includegraphics[width=1\linewidth]{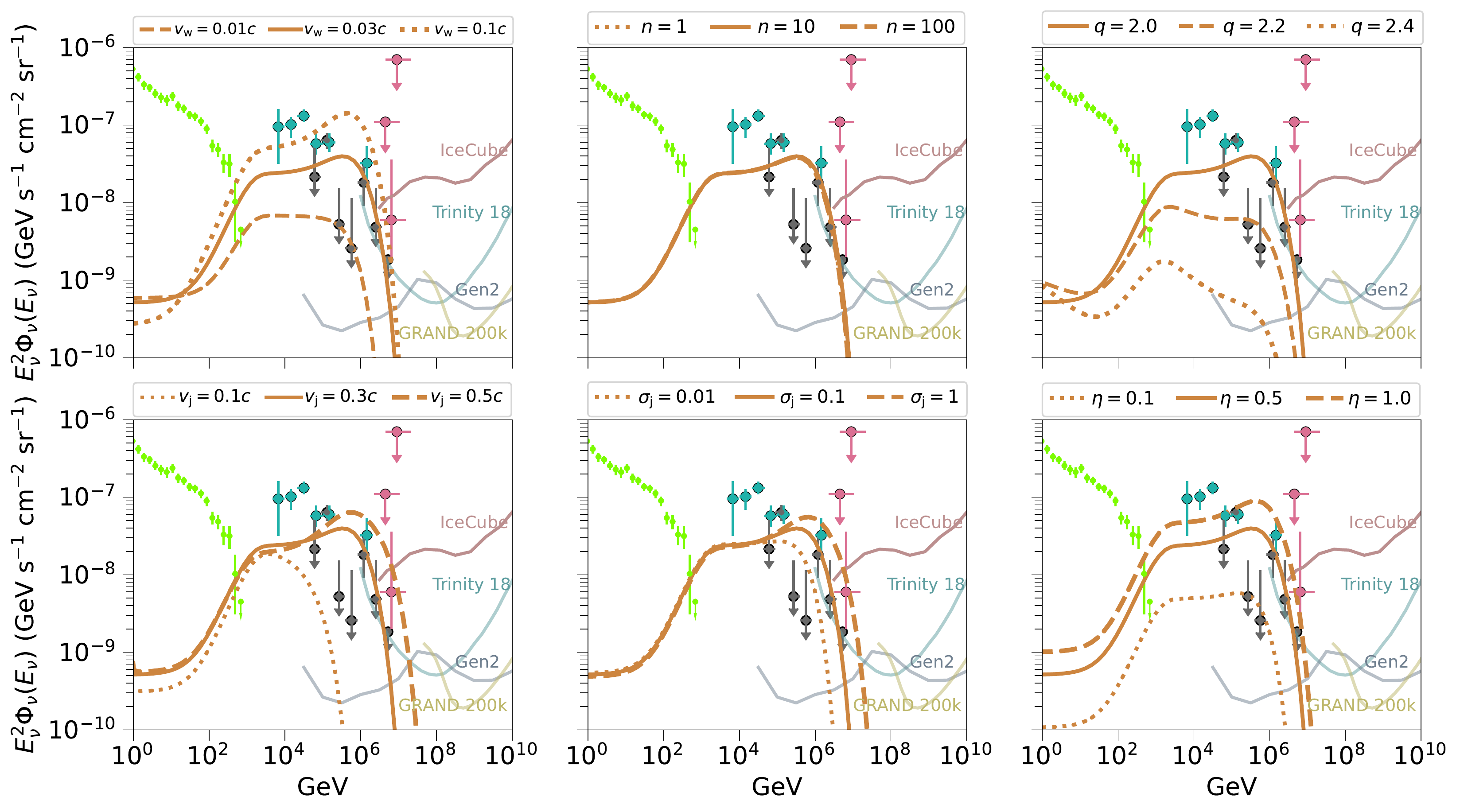}
\caption{Neutrino flux models for a range of hypernebula's physical parameters. The fiducial model is represented by solid brown curve in all the panels (but with $\dot{m}=10^7$; see also \S\ref{sec:engine_properties}). The dashed and dotted curves in each panel show how altering one parameter (that is mentioned above each panel) changes the observed flux about the fiducial model. See Fig.~\ref{fig:neutrino_flux} for a description of the other labels and markers in each panel.}
\label{fig:parameter_exploration}
\end{figure}

Fig.~\ref{fig:parameter_exploration} shows the influence of each relevant physical parameter of hypernebula on the model neutrino flux, for $\dot{m}=10^7$. As mentioned in \S\ref{subsec:neutrino_flux}, Fig.~\ref{fig:parameter_exploration} demonstrates the robustness of the low-energy turnovers for a wide range of hypernebula parameter space. The neutrino flux due to hadronuclear interactions (${\rm few}\times10^{-10}\,{\rm GeV\,s^{-1}\,cm^{-2}\,sr^{-1}}$ at $\lesssim10^2$\,GeV) is not strongly affected by any parameter except for $\eta$, which proportionately scales the normalization of the model flux at all energies as it changes only the JTS power (panel [f]).  As a result, cases with $\eta\sim1$ might not only overproduce the PeV neutrino flux, but also the diffuse gamma-ray flux. When $v_{\rm w}\ll v_{\rm j}$, an increase in $v_{\rm w}$ increases the shell radius and therefore the shock luminosity, but a further increase ($v_{\rm w}\lesssim v_{\rm j}$) would mean a slower JTS (i.e., $v_{\rm j}-v_{\rm fs}$), and a lower $n_{\rm p}$ (Eq.~\ref{eq:n_p}), which will sharply decrease the (photohadronic contribution to the) neutrino flux, and $E_{\rm max}$ (panel [a]). An increase in $v_{\rm j}$, on the other hand, monotonically increases $E_{\rm max}$, and the overall flux normalization---primarily by boosting the neutrino production through thermal photohadronic interactions (panel [d]). On the other hand, $\sigma_{\rm j}$ solely affects the high-energy cut-off (panels [b] and [e]). An increase in the proton spectral index $q$ steepens the neutrino spectrum by decreasing the thermal photohadronic component (hump at $\sim$1\,PeV) more than the nonthermal photohadronic component (hump at $\sim$1\,TeV) of the neutrino spectrum (panel [c]). While in principle, an increase in the upstream density increases the neutrino flux and high energy cutoff, it has the least influence on the overall flux ($B_{\rm jts}\propto n^{1/5}$; see Eq.~\ref{eq:B_jts}). The parameter exploration in Fig.~\ref{fig:parameter_exploration} suggests that hypernebulae---with $v_{\rm w}/c>0.1, v_{\rm j}/c\sim0.5, n\gtrsim100\,{\rm cm^{-3}}$, and $ \sigma_{\rm j}\gtrsim1$---could be promising candidates for $\gtrsim$100\,PeV UHE neutrinos that could be detected by future facilities such as IceCube-Gen2 \citep{IceCube-Gen2+14, Aartsen+21}, Trinity \citep{Nepomuk_Otte_19}, perhaps GRAND \citep{Alvarez-Muniz+20}.

\bibliographystyle{aasjournal}
\bibliography{ApJ}

\end{document}

%% file: aas_macros.tex
\def\jref@jnl#1{{\rm#1}}

\def\actaa{\jref@jnl{Acta Astron.}}      
\def\aj{\jref@jnl{AJ}}                   
\def\araa{\jref@jnl{ARA\&A}}             
\def\apj{\jref@jnl{ApJ}}                 
\def\apjl{\jref@jnl{ApJ}}                
\def\apjs{\jref@jnl{ApJS}}               
\def\ao{\jref@jnl{Appl.~Opt.}}           
\def\apss{\jref@jnl{Ap\&SS}}             
\def\aap{\jref@jnl{A\&A}}                
\def\aapr{\jref@jnl{A\&A~Rev.}}          
\def\aaps{\jref@jnl{A\&AS}}              
\def\azh{\jref@jnl{AZh}}                 
\def\baas{\jref@jnl{BAAS}}               
\def\jrasc{\jref@jnl{JRASC}}             
\def\jcap{\jref@jnl{JCAP}}               
\def\memras{\jref@jnl{MmRAS}}            
\def\mnras{\jref@jnl{MNRAS}}             
\def\na{\jref@jnl{New Astronomy}}        
\def\pra{\jref@jnl{Phys.~Rev.~A}}        
\def\prb{\jref@jnl{Phys.~Rev.~B}}        
\def\prc{\jref@jnl{Phys.~Rev.~C}}        
\def\prd{\jref@jnl{Phys.~Rev.~D}}        
\def\pre{\jref@jnl{Phys.~Rev.~E}}        
\def\prl{\jref@jnl{Phys.~Rev.~Lett.}}    
\def\pasa{\jref@jnl{PASA}}               
\def\pasp{\jref@jnl{PASP}}               
\def\pasj{\jref@jnl{PASJ}}               
\def\qjras{\jref@jnl{QJRAS}}             
\def\rmxaa{\jref@jnl{Rev.Mex.AA}}       
\def\skytel{\jref@jnl{S\&T}}             
\def\solphys{\jref@jnl{Sol.~Phys.}}      
\def\sovast{\jref@jnl{Soviet~Ast.}}      
\def\ssr{\jref@jnl{Space~Sci.~Rev.}}     
\def\zap{\jref@jnl{ZAp}}                 
\def\nar{\jref@jnl{NewAR}}               
\def\nat{\jref@jnl{Nature}}              
\def\iaucirc{\jref@jnl{IAU~Circ.}}       
\def\aplett{\jref@jnl{Astrophys.~Lett.}} 
\def\apspr{\jref@jnl{Astrophys.~Space~Phys.~Res.}}
\def\bain{\jref@jnl{Bull.~Astron.~Inst.~Netherlands}} 
\def\fcp{\jref@jnl{Fund.~Cosmic~Phys.}}  
\def\gca{\jref@jnl{Geochim.~Cosmochim.~Acta}}   
\def\grl{\jref@jnl{Geophys.~Res.~Lett.}} 
\def\jcap{\jref@jnl{JCAP}}      %
\def\jcp{\jref@jnl{J.~Chem.~Phys.}}      
\def\jgr{\jref@jnl{J.~Geophys.~Res.}}    
\def\jqsrt{\jref@jnl{J.~Quant.~Spec.~Radiat.~Transf.}}
\def\memsai{\jref@jnl{Mem.~Soc.~Astron.~Italiana}}
\def\nphysa{\jref@jnl{Nucl.~Phys.~A}}   
\def\physrep{\jref@jnl{Phys.~Rep.}}   
\def\physscr{\jref@jnl{Phys.~Scr}}   
\def\planss{\jref@jnl{Planet.~Space~Sci.}}   
\def\procspie{\jref@jnl{Proc.~SPIE}}   